\documentstyle[epsfig]{mn}

\newcommand{\bn}{{\mathbf{\nabla}}}
\newcommand{\aap}{A\&A}
\newcommand{\aj}{AJ}
\newcommand{\apj}{ApJ}
\newcommand{\apjl}{ApJ}

\newcommand{\araa}{ARA\&A}
\newcommand{\mnras}{\mbox{MNRAS}}

\begin{document}

\title [Clues on Regularity of Ellipticals] {Clues on Regularity in the Structure
and Kinematics of Elliptical Galaxies from Self-consistent Hydrodynamical Simulations:
 the Dynamical Fundamental Plane}
\author[O\~norbe et al.]
{J. O\~norbe$^1$, R. Dom\'{\i}nguez-Tenreiro$^1$, A. S\'aiz$^1$\thanks{Current address:
Dept.\ of Physics, Mahidol University, Bangkok 10400, Thailand},  
  H. Artal$^1$and A. Serna$^2$\\
  $^1$Departamento de F\'{\i}sica Te\'orica, C-XI. Universidad
  Aut\'onoma de Madrid, Madrid, E-28049, Spain\\
$^2$Departamento de F\'{\i}sica y A.C., Universidad Miguel Hern\'andez, Elche,
Spain
 }

\maketitle
\begin{abstract}
We have analysed the parameters characterising the mass and velocity
distributions of two samples of relaxed elliptical-like-objects (ELOs)
identified, at $z=0$, in a set of self-consistent hydrodynamical simulations
operating in the context of a concordance cosmological model.
ELOs have a prominent, non-rotating,
dynamically relaxed stellar spheroidal component,
with very low cold gas content, and sizes of no more than $\sim $ 10 - 40  kpc
(ELO or baryonic object scale),
embedded in a massive halo of dark matter typically
ten  times larger in size (halo scale).
They have  also an extended halo of hot diffuse gas.
The parameters characterising the mass, size and velocity dispersion 
 both at the baryonic object  and at the halo scales have been measured
 in the ELOs of each sample.
At the halo scale they have been found to satisfy virial relations;
at the scale of the baryonic object the
(logarithms of the) ELO stellar masses,
projected stellar half-mass radii, and stellar central
l.o.s. velocity dispersions define a flattened
ellipsoid close to a plane (the intrinsic dynamical plane, IDP),
tilted relative to the virial one, whose observational manifestation
is the observed FP.
Otherwise,  IDPs are  not homogeneously
populated, but ELOs, as well as elliptical (E) galaxies in the FP,
occupy only a particular region defined by the range of their masses.
The ELO samples have been found to show systematic trends with the mass
scale in both, the relative content and the relative distributions of the
baryonic and the dark mass ELO components, 
so that homology is broken in the spatial mass distribution
(resulting in the IDP tilt),
but ELOs are still a two-parameter family where the two
parameters are correlated (causing its non-homogeneous population).
The physical origin of
these trends presumably lies in the systematic decrease,
with increasing ELO mass,
of the relative amount of
dissipation experienced by the baryonic mass component along ELO
stellar mass assembly.
ELOs also show kinematical segregation,
but it does not appreciably change
with the mass scale.

The non-homogeneous population of IDPs
explains the role played by $M_{\rm vir}$ to
 determine the correlations among intrinsic parameters.
 In this paper  we also show that 
 the central stellar line-of-sight velocity dispersion
 of ELOs, $\sigma_{\rm los, 0}^{\rm star}$,
 is a fair empirical estimator of $M_{\rm vir}$,
 and this explains the central role
 played by $\sigma_{\rm los,0}$ at determining the observational
 correlations.

\end{abstract}


\section{Introduction}
\label{Intro}

Understanding how local galaxies of  different Hubble types we observe
to-day have formed is one of the most challenging open problems in cosmology.
Among the different galaxy families, elliptical (Es) are
 the easiest to study
and those that  show the most precise regularities in their
empirical properties, 
some times in the form of tight correlations
among their observable parameters.
The interest of these regularities lies in that 
they could encode a lot of relevant informations on the physical processes
  underlying E formation and evolution.
  
The Sloan digital sky survey (SDSS, York et al. 2000)
has substantially improved the statistics on  E
 samples. The sample selected by Bernardi et al. (2003a),
 using morphological and spectral criteria, contains 9000
 Es  to date in the redshift range $0.01 \le z \le 0.3$
 and in every environment from
 voids to groups to rich clusters.
  Analyses of their structural and dynamical parameters have shown that
  the distributions of their luminosities $L$,
  radii at half projected light $R_{\rm e}^{\rm light}$, and
  central line-of-sight  velocity dispersions $\sigma_{\rm los, 0}$
    (Bernardi et al. 2003b, 2003c),
    are approximately gaussian at any $z$.
Moreover, a maximum likelihood analysis indicates that
the pairs of parameters 
$\sigma_{\rm los, 0}$---$L$ and 
$R_{\rm e}^{\rm light}$---$L$,  or their combinations,
such as the mass-to-luminosity
ratio within the effective radii  $M_{\rm e}/L$ and
$L$
(where $M_{\rm e}$ is the dynamical mass defined as
$M_{\rm e} = 2 R_{\rm e}^{\rm light} \sigma_{\rm los, 0}^{2}/G$),
show correlations consistent  with those previously
established in literature,
obtained from individual galaxy spectra
of smaller samples, such as the Faber-Jackson
relation (1976);
the $D_n$---$\sigma_{\rm los, 0}$ relation (Dressler et al. 1987); and
 the surface brightness ---$R_{\rm e}$ relation (Kormendy 1977,
Kormendy \& Djorjovski 1989), among others.
Furthermore,  early-type galaxies in the SDSS have been found to have
roughly constant stellar-mass-to-light ratios (Kauffmann et al. 2003a, 2003b;
Padmanabhan et al. 2004).

 The correlations involving two variables out of the three needed
 to fully characterise the structure and dynamics of an E galaxy,
 are projections of the so-called Fundamental Plane relation
 (FP, Djorgovski \& Davis 1987; Dressler et
 al. 1987a; Faber et al. 1987; Kormendy \& Djorgovski 1989),
 involving the three or some or their combinations.
 The FP relation can be written as

 \begin{equation}
 \log_{10} R_{\rm e}^{\rm light} = a \log_{10} \sigma_{\rm los, 0} + b \log_{10} <I^{\rm light}>_{\rm e} + c.
 \label{FP}
 \end{equation}

 where $<I^{\rm light}>_{\rm e}$ is the mean surface brightness within
 $R_{\rm e}^{\rm light}$.
 The values of the FP coefficients for the SDSS E  sample are $a \simeq 1.5$,
 similar in the four SDSS bands,
 $b \simeq -0.77$, and $c \simeq -8.7$ (see their exact values in   Bernardi et al. 2003c,
 Table 2) with a small scatter.
 These SDSS results confirm previous ones,
 either in the
 optical (Lucey, Bower \& Ellis 1991; de Carvalho \& Djorgovski 1992;
 Bender, Burstein \& Faber 1992; J{\o}rgensen et al. 1993;
 Prugniel \& Simien 1996;
 J{\o}rgensen et al. 1996)
 or in the near-IR wavelengths (Recillas-Cruz et al. 1990, 1991;
 Pahre, Djorgovski \& de Carvalho 1995;
 Mobasher et al. 1999), even if the published values of $a$ show larger
 values in the $K$-band
 than  at shorter wavelengths
 (see, for example, Pahre, de Carvalho \& Djorgovski
 1998).
 
   The existence of the FP and its small scatter has the important
   implication that it provides us with a strong constraint
   when studying elliptical galaxy formation and evolution
   (Bender, Burstein \& Faber 1993; Guzm\'an, Lucey \& Bower 1993;
   Renzini \& Ciotti 1993).
The physical origin of the FP is not yet clear,
but it must be a consequence of the physical processes
responsible for galaxy assembly.
These processes built up early type galaxies as dynamically hot systems
whose configuration in phase space are close to equilibrium.
Taking an elliptical galaxy as a system in equilibrium,
the virial theorem

\begin{equation}
M_{\rm vir} = c_{\rm F} (\sigma_{3, h}^{\rm tot})^{2} r_{\rm e, h}^{\rm tot}/ G ,
\label{Virial}
\end{equation}

where  $M_{\rm vir}$ is its virial mass, 
$\sigma_{\rm 3, h}^{\rm tot}$ is the average 3-dimensional velocity dispersion
of the whole elliptical, including both dark and baryonic matter,
$r_{\rm e, h}^{\rm tot}$ the dynamical half-radius or radius enclosing
half the total mass of the system, and $c_{\rm F}$ a form factor
of order unity, would imply a relation similar to Eq. \ref{FP}
with $a=2$ and $b=-1$ (known as the virial plane, Faber et al. 1987), 
should the 
dynamical mass-to-light ratios, $M_{\rm vir}/L$, and the 
mass structure coefficients

\begin{equation}
c_{\rm M}^{\rm vir} = {G M_{\rm vir} \over 3 \sigma_{\rm los, 0}^{2} R_{\rm e}^{\rm light}}, 
\label{CMvir}
\end{equation}

be independent of the E luminosity or mass scale.
The observational results described above mean that the FP is
{\it tilted} relative to the virial plane, and, consequently that
either $M_{\rm vir}/L$ or $c_{\rm M}^{\rm vir}$, or both, do depend
on the E luminosity.

Different authors interpret the tilt of the FP relative to the
virial relation as caused by different misassumptions that we comment
briefly (note that we can write
 $M_{\rm vir}/L =  M^{\rm star}/L \times M_{\rm vir}/M^{\rm star}$,
where $M^{\rm star}$ is the stellar mass of the elliptical galaxy).

1.1) A first possibility
is that the tilt is due to
systematic changes of stellar age and metallicity with galaxy mass, or, even,
to changes of the slope of the stellar initial mass function (hereafter, IMF)
with galaxy mass, resulting in systematic changes in the
{\it stellar}-mass-to-light ratios, $M^{\rm star}/L$, with mass
or luminosity
 (Zepf \& Silk 1996; Pahre at al. 1998; Mobasher et al. 1999).
 But these effects  could explain at most only
 $\sim$ one third of the $\beta \neq 0$  value
 in the $B$-band (Tinsley 1978; Dressler et al. 1987;
  Prugniel \& Simien 1996; see also
  Renzini \& Ciotti 1993;
  Trujillo, Burkert \& Bell 2004).
Furthermore,  early-type galaxies in the SDSS have been found to have
roughly constant stellar-mass-to-light ratios (Kauffmann et al. 2003a, 2003b).
Anyhow, the presence of a tilt in the $K$-band FP,
where population effects are no important, indicates
that it is very difficult that the tilt
is caused by stellar physics processes alone,
as Bender et al. (1992), Renzini \& Ciotti (1993),
Guzm\'an et al. (1993),
Pahre et al. (1998),
among other authors, have suggested.

1.2) A second possibility is that $M_{\rm vir}/L$ changes systematically
with the mass scale because the total dark-to-visible mass ratio,
$M_{\rm vir}/M^{\rm star}$ changes
(see, for example,
Renzini \& Ciotti 1993; Pahre et al. 1998;
Ciotti, Lanzoni \& Renzini 1996; Padmanabhan et al. 2004).

Otherwise, a dependence of $c_{\rm M}^{\rm vir}$ on the mass
scale could be caused by systematic differences in:

2.1) the dark  versus bright matter spatial distribution,

2.2) the kinematical segregation,
the rotational  support and/or velocity
dispersion anisotropy in the stellar
component  (dynamical non-homology),

2.3) systematic  projection or other geometrical effects.

Taking into account these effects in the FP tilt demands
modelling the galaxy mass and velocity three-dimensional distributions
and comparing the outputs with high quality data.

Bender et al. (1992) considered  effects iii) and iv);
Ciotti et al. (1996) explore ii) - iv) and conclude that
an systematic increase in the dark matter content with mass, or differences
in its distribution, as well as a dependence of the S\'ersic (1968)
shape parameter for the luminosity profiles
with mass, may by themselves formally
produce the tilt; Padmanabhan et al. (2004) find evidence of
effect ii) in SDSS data.
Other authors have also shown that allowing for broken homology,
either dynamical (Busarello et al. 1997), in the luminosity
profiles (Trujillo et al. 2004), or both
(Prugniel \& Simien 1997;
Graham \& Colless 1997; Pahre et al. 1998),
brings the observed FP closer to Eq. (3).

Observational methods suffer from some drawbacks
to study in deph the  physical origin of the FP tilt. For example,
 a drawback
is the impossibility
to get accurate measurements of the elliptical
three-dimensional  mass distributions
(either dark, stellar or gaseous) 
 and  another is that  the intrinsic 3D velocity distribution 
of galaxies is  severely limited by projection.
Only 
 the line-of-sight velocity distributions 
can be inferred from galaxy spectra.
And, so, the interpretation  of observational data
is not always straightforward.
To complement the informations provided by data and circumvent these drawbacks,
analytical modelling is largely used in literature
(Kronawitter et al. 2000; Gerhard et al. 2001;
Romanowsky \& Kochanek 2001;
Borriello et al. 2003; Padmanabhan et al. 2004;
Mamon \& Lockas 2005a, 2005b).
These methods give very interesting insights into
mass and velocity distributions, as well as the physical
processes causing them, but are somewhat
limited by symmetry  considerations and other necessary  simplifying
hypotheses.
These difficulties and limitations could be circumvented should we have
at our disposal complete  informations on the phase-space
of the galaxy constituents.
This is not possible through observations,
but can be attained, at least in a virtual sense, through
numerical simulations.

Capelato, de Carvalho \& Carlberg (1995)  first addressed the origin of the FP through numerical
simulations. By analyzing the remnants of the dissipationless mergers of two
equal-mass one-component King models, and varying their relative orbital energy
and angular momentum, they show that the mergers of objects in the FP
produces a new objects in the FP. This result was extended by Dantas et al.
(2003), who used one- and two-component Hernquist models as progenitors,
 Goz\'alez-Garc{\'\i}a \& van Albada (2003), based on Jaffe (1983)
models and by Boylan-Kolchin et al. (2005), who used Hernquist+NFW models. 
Nipoti, Londrillo \& Ciotti (2003) show, in turn, that the
FP is well reproduced by dissipationless hierarchical equal-mass  merging
of one- and two-component galaxy models, and by accretion with substantial
angular momentum, with the merging zeroth-order generation placed at the
FP itself. They also found that both the Faber-Jackson and the Kormendy
relations are not reproduced by the simulations, and conclude that
dissipation must be a basic ingredient in elliptical formation.
In agreement with this conclusion,
Dantas et al. (2002) and Dantas et al. (2003) have shown that the
end products of  dissipationless collapse generally do not follow
a FP-like correlation.
Bekki (1998) first considered the role of dissipation in elliptical
formation through pre-prepared simulations. He adopts the merger hypothesis
(i.e.., ellipticals form by the mergers of two equal-mass gas-rich
spirals) and he focuses on the role of the timescale for star formation
in determining the structural and kinematical properties of the
merger remnants. He concludes that the slope of the FP reflects the
difference in the amount of dissipation the merger end products
have experienced according with their luminosity (or mass).
Recently, Robertson et al. (2006) have confirmed
this conclusion on the role of dissipative dynamics
to shape the FP, again  through  pre-prepared
mergers of disk galaxies.

 In this paper we go a step further
 and we analyse the tilt of the FP in samples of virtual ellipticals
 formed in a cosmological context through self-consistent
 hydrodynamical simulations.
 This numerical approach provide
 a convenient method to work out 
the clues of regularity and systematics of elliptical galaxies,
and to find out their links with the processes involved
in galaxy assembly in a cosmological context.
 The point important for our present purposes is that they {\it directly}  
provide  with complete 6-dimensional
phase-space informations on each constituent particle sampling a
given galaxy-like object formed in the simulation,
that is, they give directly the mass and velocity  distributions
of dark matter, gas and stars of each object.

Taking advantage of these possibilities, we
have analysed ten self-consistent hydrodynamical   
simulations run in the framework of
a flat $\Lambda$CDM cosmological model, characterised
by cosmological parameters consistent with their
last determinations (Spergel et al. 2006). 
Galaxy-like objects of different morphologies form in these
 simulations at $z=0$: irregulars, disc-like objects, S0-like objects and
 elliptical-like objects (hereafter,
ELOs). ELOs have been identified as those objects having a prominent
dynamically relaxed, roughly non-rotating stellar spheroidal component,
with no extended disks and very low cold gas content;
the  stellar component
has typical sizes of no more than $\sim $ 10 - 40  kpc,
it dominates the mass density at these scales (hereafter, {\it ELO or baryonic object  scale})
and it
is embedded in a halo of dark matter typically ten  times larger in size
(hereafter, {\it halo scale}). 
In a forthcoming paper (O\~norbe et al., in preparation)
we report on an analysis of the mass and velocity 
distributions of the different ELO components: dark matter,
stars, cold gas and hot gas. 
In this paper we focus on the quantitative characterisation of these mass
and velocity distributions
through their corresponding parameters, both at the ELO and at the halo scales.
At the baryonic object scale,
to characterise the structural and dynamical properties of ELOs, we will
describe their three dimensional  distributions
of mass and velocity
through the three intrinsic (that is, three-dimensional) parameters
(the stellar mass at the baryonic object  scale, $M_{\rm bo}^{\rm star}$,
the stellar  half-mass radii
at the baryonic object  scale, $r_{\rm e, bo}^{\rm star}$,
defined as those radii enclosing half the $M_{\rm bo}^{\rm star}$ mass,
and the mean square
velocity for stars, $\sigma_{\rm 3, bo}^{\rm star}$)
whose observational projected counterparts
(the luminosity $L$, effective projected size  $R_{\rm e}^{\rm light}$, and the stellar
central l.o.s. velocity dispersion, $\sigma_{\rm los, 0}$) enter the
definition of the observed FP (Eq. \ref{FP}).
To help the reader, in Table~\ref{tabnombres} we give a list 
of the parameter names and symbols and in Table~\ref{tabnombres2} a list of the profiles and ratios.
These intrinsic parameters have been measured on ELOs, and their 
correlations have been looked for, and more specifically, the lower--dimensional
 regions (i.e., {\it dynamical planes}) they fill in the
three dimensional space of the $M_{\rm bo}^{\rm star}$,
$r_{\rm e, bo}^{\rm star}$
and $\sigma_{\rm 3, bo}^{\rm star}$ parameters, because the observational
manifestation of these dynamical planes is the FP relation.
These informations, combined with that on  the mass,
size and velocity dispersion parameters at the halo scale, 
allows us to test whether or not
  the $c_{\rm M}^{\rm vir}$ coefficients and the 
  $M_{\rm vir}/M^{\rm star}$ ratios do  systematically
depend on the mass scale, so that 
the tilt and the scatter of the observed FP can be explained
in terms of the  regularities in the structural and dynamical
properties of ELOs formed in self-consistent
hydrodynamical simulations.

Fully-consistent gravo-hydrodynamical simulations as a method
to study E assembly has already proven to be useful. 
An analysis  of ELO  structural and kinematical properties
that can be constrained from observations
(i.e., stellar masses,  projected half-mass  radii,
central line-of-sight velocity dispersions), has 
shown that they have counterparts in the local Universe
as far as these properties are concerned
(S\'aiz 2003; S\'aiz, Dom\'{\i}nguez-Tenreiro 
\& Serna 2004, hereafter SDTS04), including the FP relation
and some clues about its physical origin
(O\~norbe et al. 2005), and its lack of dynamical evolution
(Dom\'{\i}nguez-Tenreiro et al. 2006, hereafter DTal06).
Also, ELO stellar populations have age distributions
showing similar trends as those inferred from observations
(Dom\'{\i}nguez-Tenreiro, S\'aiz \& Serna 2004, hereafter DSS04).

The paper is organised as follows:
in $\S$2 we briefly describe the simulations,
the ELO samples and their generic properties.
The ELO size and mass scales and their relations are analysed in $\S$3.
 $\S$4 is devoted to kinematics and in $\S$5
 we report on the intrinsic dynamical plane of ELOs
 and its comparison with the observed Fundamental Plane,
 and 
 we analyse its physical origin. 
 Finally, in $\S$6 we summarise our results and discuss them 
 in the context of theoretical results on halo structure
 and dissipation of the gaseous component.

\section{The Simulations and the ELO Samples}
\label{simula}

We have analysed ELOs identified in ten
self-consistent cosmological simulations run  
in the framework of the same
global  flat $\Lambda$CDM cosmological model,
with $h=0.65$,
$\Omega_{\rm m} = 0.35$, $\Omega_{\rm b} = 0.06$.
The normalisation parameter has been taken slightly high,
$\sigma_8 = 1.18$, as compared with the average fluctuations
of 2dFGRS or SDSS  galaxies
(Lahav et al. 2002, Tegmark et al. 2004) or recent results from WMAP
(Spergel et al. 2006) to mimic an active region of the Universe
(Evrard, Silk \& Szalay 1990).

We have used a lagrangian code,
(\textsc{DEVA}, Serna, Dom\'{\i}nguez-Tenreiro \& S\'aiz, 2003),
particularly designed to study
 galaxy assembly in a cosmological context.
  Gravity is computed through an AP3M-like  method, based
  on Couchman (1991).
  Hydrodynamics is  computed through a SPH technique where 
  special attention has been paid to make 
the implementation of conservation laws (energy, entropy
  and angular momentum) as accurate as possible
   (see Serna et al. 2003 for details,
   in particular for  a discussion on the observational implications of
   violating some conservation laws).
 Entropy
 conservation  is assured by taking into consideration the space variation
 of the smoothing length (i.e., the so-called $\bn h$ terms).
   Time steps are individual for particles
  (to save CPU time, allowing a good time resolution),
 as well as masses. Time
		    integration uses a PEC scheme.
In any run, an homogeneously sampled
periodic box of 10~Mpc side  has been employed
and  64$^3$
dark matter and 64$^3$ gas particles,
with a mass of $1.29 \times
10^8$ and $2.67 \times 10^7 $M$_{\odot}$, respectively, have
been used.
The gravitational softening used was $\epsilon = 2.3$ kpc.
The cooling function is that from Tucker (1975)  and Bond et al. (1984)
for an optically thin primordial mixture of H and He ($X=0.76$,
$Y=0.24$) in collisional equilibrium and in absence of any
significant background radiation field
with a primordial gas composition.
Each  of the ten simulations started at a redshift $z_{\rm in} = 20$.

Star formation (SF) processes have been included through a simple 
phenomenological parametrisation, as that first used by Katz (1992,
see also Tissera et al. 1997; S\'aiz 2003 and Serna et al. 2003 for
details)
  that transforms   cold locally-collapsing gas
  at the scales the code resolves,
  denser than a threshold density,
  $\rho_{\rm thres} $,
  into stars at a rate
  $d\rho_{\rm star}/dt = c_{\ast}  \rho_{\rm gas}/ t_{\rm g}$,
  where  $t_g$ is a characteristic time-scale chosen
  to be equal to the maximum of the local gas-dynamical time,
  $t_{dyn} = (4\pi G\rho_{\rm gas})^{- 1/2}$,
  and the local cooling time;   $c_{\ast}$ is the average
  star formation efficiency at resolution  $\epsilon $ scales, i.e.,
  the empirical Kennicutt-Schmidt law (Kennicutt 1998).
It is worth noting that, in the
context of the new sequential multi-scale SF scenarios
(V\'azquez-Semadeni 2004a, 2004b; Ballesteros-Paredes et al. 2006
and references therein),
 it has been argued that
 this law, and particularly so the low $c_{\ast}$ values
 inferred from observations, can be explained as a result of
 SF processes acting on dense molecular cloud core scales
 when conveniently averaged on disc scales
 (Elmegreen 2002; Sarson et al.\ 2004, see below).
Supernova feedback effects or  energy inputs other than gravitational
have not been  {\it explicitly} included in these simulations.
We note that the  role of discrete stellar energy sources at the
 scales resolved in this work is not yet clear,
  as some authors argue that
   stellar energy releases drive the structuring of the gas density 
    locally at sub-kpc scales (Elmegreen 2002).
      In fact, some MHD simulations of self-regulating SNII
        heating in the ISM at scales $<$ 250 pc
	  (Sarson et al.\ 2004), indicate that
	    this process produces a Kennicutt-Schmidt-like law on
	      average. If this were the case, the Kennicutt-Schmidt law
	        implemented in our code would already  {\it implicitly}
		  account for the effects
		    stellar self-regulation has on the scales our code resolves,
		      and our ignorance on sub-kpc scale processes would be
contained in the particular values of $\rho_{\rm thres} $
and $c_{\ast}$.

Five out of the ten simulations
(the SF-A type simulations)
share the SF parameters ($\rho_{\rm thres} = 6 \times 10^{-25} $ gr cm$^{-3}$,
$c_*$ = 0.3) and differ in the seed  used to build up the initial conditions.
To test the role of SF parameterisation,
the same initial conditions
have been run with different SF parameters
($\rho_{\rm thres}$ =  $1.8 \times 10^{-24} $ gr cm$^{-3}$,
$c_*$ = 0.1) making SF  more difficult, contributing another set of five
simulations  (hereafter, the SF-B type simulations).

ELOs have been identified in the simulations as those
galaxy-like objects  at $z = 0$ having a prominent, non-rotating,
dynamically relaxed spheroidal   component made out
of stars, with no extended discs and very low cold gas content.
It turns out that, at $z=0$,
26 (17) objects out of the more massive  formed in
SF-A (SF-B) type simulations fulfil this identification criterion,
forming two  samples
(the SF-A and SF-B ELO samples)
partially analysed
in SDTS04, in DSS04
and in DTal06.
In O\~norbe et al. (2005) it is shown that both samples
satisfy dynamical FP relations.		    
ELOs in the SF-B sample tend to be of later type than their
corresponding SF-A counterparts because forming stars becomes more difficult;
 this is why many of
  the SF-B sample counterparts of the less massive ELOs
   in SF-A sample do not satisfy the selection criteria, and the SF-B sample has
    a lower number of ELOs that the SF-A sample.
	       
A visual examination of ELOs indicates that 
the stellar component is embedded in
a dark matter halo, contributing an important fraction
of the mass at distances from the ELO centre larger than
$\sim 15$ kpc on average. ELOs have also
a hot, extended, X-ray emitting  halo of diffuse gas 
(S\'aiz, Dom\'{\i}nguez-Tenreiro \& Serna 2003).
Stellar and dark matter particles constitute a dynamically hot
component with an important velocity dispersion, and, except
in the very central regions, a positive anisotropy.
    In  Table~\ref{tabhaloscale} and Table~\ref{tabetloscale}
    different data on ELOs in the samples are given.
    In these Tables, the criterion introduced in S\'aiz et al. 2001,
    ELOs have been labelled by a three--digit code formed by rounding
    their $x$, $y$ and $z$ coordinates in units where
    the simulated box has, at any redshift, a length of one.
    The number of particles of each species sampling each ELO in the sample
    can be easily determined from these Tables and the values of
    the masses of dark and baryonic particles.
The   spin parameters of the ELO samples have
an average value of $\bar{\lambda} = 0.033$.
ELO  mass function is consistent
with that of a small group environment
(Cuesta-Bolao \& Serna, private communication).

The simulations show the physical  patterns of
ELO mass assembly and star formation (SF) rate histories
(see  DSS04 and DTal06 for more details).
ELOs form  out of  the mass elements that at high $z$ are
enclosed by   overdense regions whose mass scale (total mass they
enclose) is of the order of an E galaxy total (i.e., including
its halo) mass.
Analytical models, as well as N-body simulations indicate that two different
phases operate  along  halo mass assembly: first, a violent fast one,
where the mass aggregation rates are high,
and then, a  slower one, with lower mass aggregation rates
(Wechsler et al. 2002; Zhao et al. 2003; Salvador-Sol\'e, Manrique, {\&} Solanes 2005).
Our hydrodynamical simulations indicate
that the fast phase occurs through a
multiclump collapse (see Thomas, Greggio \& Bender 1999)
ensuing turnaround of the overdense regions,
and it is characterised by the fast head-on fusions experienced by
the nodes of the cellular structure these regions enclose,
resulting in strong shocks and  high cooling rates of their gaseous component,
and, at the same time, in strong and very fast star formation bursts 
that transform most of the available cold  gas into stars.
Consequently, most of the dissipation involved in the mass assembly of
a given ELO occurs  in this violent early phase at high $z$.
The slow phase comes after the multiclump collapse.
In this phase, the halo mass aggregation rate is low and the mass
increments result from major mergers, minor mergers or continuous
accretion. Our cosmological simulations show that
the  fusion rates are generally low,
and that a strong star formation burst
and dissipation  follow a major merger
  only if enough gas is still available after the early violent phase.
    This is very unlikely in any case, and it becomes more and more
      unlikely as the ELO mass increases (see DSS04).
        And so,
these mergers imply only a modest amount of energy dissipation
or star formation, but they play an important role in this slow phase:
an $\sim $ 50\% of ELOs in the sample have experienced a major merger
event at $ 2 < z < 0$, that result in the increase of
the ELO mass content,
 size, and stellar mean square velocity.
So, our simulations indicate that most of the stars of to-day ellipticals
formed at high redshifts, while they are assembled later on
(see de Lucia et al. 2006, for similar results from a semi-analytic model
of galaxy formation grafted to the {\it Millennium Simulation}).
This scenario shares 
some characteristics of previously proposed scenarios
(see discussion and references in DTal06),
but it has also significant differences, mainly
that most stars form out of cold gas that had never been shock heated
at the halo virial temperature and then formed a disk,
as the conventional recipe for galaxy formation propounds
(see discussion  in Keres et al. 2005 and references therein).

\section{Size and Mass Scales}
\label{3DVirialResults}
 
\subsection{Masses and Sizes at the Scale of the Virial Radius}
\label{MaSizes}

The virial radius describes the ELO size  {\it at the scale
of its dark matter halo}. It roughly  encloses those particles
that are bound into the self-gravitating configuration
forming a given ELO system (i.e., a dark matter halo plus
the main baryonic compact object plus  the substructures
and satellites hosted by the
dark matter halo).
The virial radii  have been calculated
using the Bryan \& Norman (1998) fitting formula, that yields, at $z=0$,
a value of $\Delta \simeq 100$ for the mean density within
$r_{\rm vir}$ in units of the critical density.
The mass  at the scale of  $r_{\rm vir}$ is the virial
mass, $M_{\rm vir}$, the total mass inside $r_{\rm vir}$
or halo total mass.
The mass scales associated to the
different constituents  considered here are\footnote{Note that we
have used superscripts to mean the different ELO constituents,
and subscripts to distinguish between   halo (h) or baryonic
object  (bo) scales, see Table~\ref{tabnombres}}:
dark matter, $M_{\rm h}^{\rm dark}$, baryons of any kind,
$M_{\rm h}^{\rm ab}$, cold baryons
(that is, cold gas particles with  $T \le 3 \times 10^{4}$K
and stellar particles),
$M_{\rm h}^{\rm cb}$, stars, $M_{\rm h}^{\rm star}$, and
hot gas (that is, gaseous particles with $T > 3 \times 10^{4}$K).
A measure of the compactness of the mass
distribution for the different ELO constituents, at the
halo scale, is given by their respective half-mass radii, or
radii enclosing half the mass of these constituents
within $r_{\rm vir}$; for example, the overall half-mass radii,
$r_{\rm e, h}^{\rm tot}$, are the radii of the sphere enclosing $M_{\rm vir}/2$,
 the stellar half-mass radii $r_{\rm e, h}^{\rm star}$
 enclose $M_{\rm h}^{\rm star}/2$ and so on.

\begin{table*}[H]

\caption[Parameter names and symbols]{Parameter names and symbols}
\label{tabnombres}

\begin{center}
\footnotesize

\begin{tabular}{lc}
\hline
\hline
 Name & Symbol \\
\hline
\hline
Observational parameters & \\
\hline
Luminosity& $L$  \\
Half projected light radius& $R_{\rm e}^{\rm light}$  \\
Central LOS velocity dispersion&  $\sigma_{\rm los, 0}$ \\
Dynamical mass & $M_{\rm e}$ \\
Mean surface brightness within $R_{\rm e}^{\rm light}$ & $<I^{\rm light}>_{\rm e}$ \\
Stellar Mass & $M^{\rm star}$ \\
Stellar-mass-to-light ratio & $\gamma^{\rm star}$ \\
\hline
Halo scale parameters & \\
\hline
Virial mass &$M_{\rm vir}$ \\
Virial radius&$r_{\rm vir}$  \\
Dark mass inside virial radius&$M_{\rm h}^{\rm dark}$ \\
Baryon mass inside virial radius&$M_{\rm h}^{\rm ab}$ \\
Cold baryon mass inside virial radius&$M_{\rm h}^{\rm cb}$ \\
Stellar mass inside virial radius&$M_{\rm h}^{\rm star}$ \\
Total half-mass radius&$r_{\rm e, h}^{\rm tot}$ \\
Cold baryon half-mass radius&$r_{\rm e, h}^{\rm cb}$ \\
Stellar half-mass radius&$r_{\rm e, h}^{\rm star}$ \\
Total 3D velocity dispersion&$\sigma_{\rm 3, h}^{\rm tot}$ \\
\hline
Baryonic-object scale parameters &  \\
\hline
Stellar mass & $M_{\rm bo}^{\rm star}$\\
Cold baryon mass & $M_{\rm bo}^{\rm cb}$ \\
Stellar half-mass radius & $r_{\rm e, bo}^{\rm star}$\\
Cold baryon half-mass radius & $r_{\rm e, bo}^{\rm cb}$ \\
Projected stellar half-mass radius& $R_{\rm e, bo}^{\rm star}$\\
Mean stellar 3D velocity dispersion & $\sigma_{\rm 3, bo}^{\rm star}$\\
Central LOS stellar velocity dispersion & $\sigma_{\rm los, 0}^{\rm star}$\\
Mean projected stellar mass density within $R_{\rm e, bo}^{\rm star}$ & $< \Sigma^{\rm star}>_{\rm e}$ \\

\hline
\hline
\end{tabular}
\end{center}
\end{table*}

\begin{table*}
  \caption[Profiles  and ratios]{Profiles and ratios}
\label{tabnombres2}

  \begin{center}
 \footnotesize

 \begin{tabular}{lclcc}
\hline
\multicolumn{2}{c}{Profiles} &
\multicolumn{3}{c}{Ratios} \\
\hline
Name & Symbol$^a$ & Ratio definition & Ratio symbol & Logarithmic slope\\
\hline
\hline
Hot baryon mass profile& $M^{\rm hb}(r)$ & $G M_{\rm vir} / (\sigma_{\rm 3, h}^{\rm tot})^2  r_{\rm e, h}^{\rm tot}$ & $c_{\rm F} $ & $\beta_{\rm F}$ \\
Circular velocity profile & $V_{\rm cir}(r)$ & $r_{\rm e, h}^{\rm tot} / r_{\rm e, bo}^{\rm star}$ & $c_{\rm rd}$ & $\beta_{\rm rd}$\\
3D  velocity dispersion profile&$\sigma_{3D}(r)$ & $r_{\rm e, bo}^{\rm star} / R_{\rm e, bo}^{\rm star}$ & $c_{\rm rp}$ & $\beta_{\rm rp}$\\
Anisotropy profile & $\beta_{\rm ani}(r)$ & $(\sigma_{\rm 3, h}^{\rm tot} / \sigma_{\rm 3, bo}^{\rm star})^2$ & $c_{\rm vd}$ & $\beta_{\rm vd}$\\
Projected mass density profile & $\Sigma(R)$ & $(\sigma_{\rm 3, bo}^{\rm star})^2 / 3 (\sigma_{\rm los, 0}^{\rm star})^2$ & $c_{\rm vpc}$ & $\beta_{\rm vpc}$ \\
Line-of-sight velocity profile &$V_{\rm los}(R)$ & ${G M_{\rm vir} / 3 (\sigma_{\rm los, 0}^{\rm star})^{2} R_{\rm e}^{\rm light}} = c_{\rm F} c_{\rm rd} c_{\rm rp} c_{\rm vd} c_{\rm vpc}$ & $c_{\rm M}^{\rm vir}$ & $\beta_{\rm M}$\\
Line-of-sight velocity dispersion profile & $\sigma_{\rm los}(R)$&&&\\
\hline
\hline

\end{tabular}
\end{center}

\medskip
(a) To specify the constituent, a superindex has been added in the text to the profile symbols
\end{table*}

\begin{figure}
  \begin{center}
    \includegraphics[width=.45\textwidth]{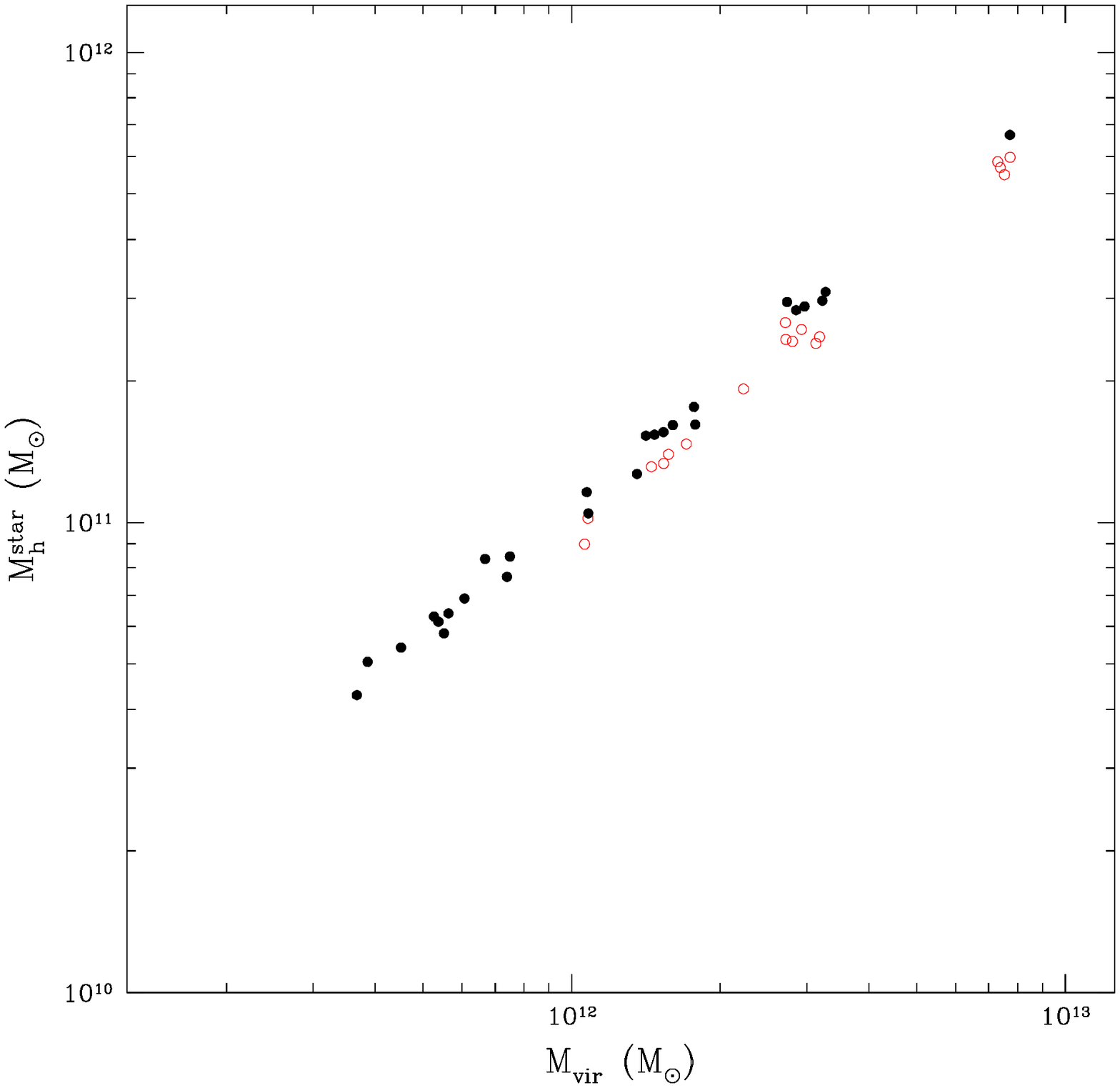}
    \caption{
      Masses at the halo scale of  stars
      versus their corresponding virial masses.
      filled black symbols: SF-A sample ELOs; open red symbols: SF-B sample ELOs
      }
    \label{MvirCorr}
  \end{center}
\end{figure}

For each of the ELOs in our sample, their virial mass and  radii
are listed in Table~\ref{tabhaloscale},
where also given are the masses  within $r_{\rm vir}$
corresponding to different constituents and some relevant
half-mass radii.
All these mass scales are strongly correlated with
$M_{\rm vir}$ as shown in Figure~\ref{MvirCorr}
for
 $M_{\rm h}^{\rm star}$.      
Note that the
virial masses of ELOs have a lower limit  of 3.7 $\times$ 10$^{11}$ M$_{\odot}$.

\begin{table*}
  \caption[Masses, sizes and mean square velocities  of ELOs at the halo scale]
  {Masses, sizes and mean square velocities  of ELOs at the halo scale
  ($z=0$)}
  \label{tabhaloscale}
 
  \begin{center}
    \footnotesize       

\begin{tabular}{lccccccccccc}
    \hline
     Run&ELO&$M_{\rm vir}$&$M_{\rm h}^{\rm dark}$& $M_{\rm h}^{\rm ab}$& $M_{\rm h}^{\rm cb}$& $M_{\rm h}^{\rm star}$&  $r_{\rm vir}$ & $r_{\rm e, h}^{\rm tot}$& $r_{\rm e, h}^{\rm cb}$ & $r_{\rm e, h}^{\rm star}$ & $\sigma_{\rm 3, h}^{\rm tot}$ \\
    \hline
    \hline
8714 & \#173 & 772.82 & 678.31 & 94.51 & 70.65& 66.69& 527.00& 222.54& 64.80& 51.63& 302.43 \\
           & \#353 & 322.18 & 285.71 & 36.47 & 31.60& 29.61& 394.00& 110.99& 18.62& 16.11& 261.18 \\
           & \#581 & 177.03 & 156.26 & 20.77 & 18.94& 17.59& 322.00& 101.54& 19.05& 16.07& 202.53 \\
           & \#296 & 153.52 & 134.38 & 19.14 & 16.79& 15.56& 308.00& 92.12& 17.09& 13.20& 201.13 \\
           & \#373 & 74.00 & 64.23 & 9.77 & 8.35& 7.66& 241.00& 66.65& 6.70& 5.79& 162.36 \\
           & \#772 & 60.71 & 52.16 & 8.55 & 7.50& 6.89& 226.00& 56.57& 6.58& 5.40& 159.75 \\
           & \#284 & 53.76 & 46.03 & 7.73 & 6.86& 6.15& 217.00& 67.69& 6.34& 4.96& 149.92 \\
    \hline
    \hline
8747 & \#288 & 285.17 & 251.80 & 33.37 & 29.83& 28.27& 378.00& 112.14& 25.87& 22.65& 241.90 \\
           & \#115 & 178.07 & 157.31 & 20.77 & 17.62& 16.14& 323.00& 95.56& 11.28& 8.24& 220.92 \\
           & \#189 & 45.15 & 38.62 & 6.53 & 5.77& 5.42& 205.00& 58.45& 4.88& 4.45& 145.62 \\
           & \#915 & 38.63 & 32.47 & 6.17 & 5.51& 5.05& 194.00& 47.74& 4.14& 3.59& 141.02 \\
    \hline
    \hline
8741 & \#011 & 273.34 & 237.46 & 35.87 & 30.12& 29.42& 373.00& 102.20& 21.83& 20.37& 246.70 \\
           & \#017 & 141.44 & 121.97 & 19.47 & 16.33& 15.29& 299.00& 89.04& 17.01& 13.93& 197.56 \\
           & \#930 & 135.71 & 118.56 & 17.15 & 13.70& 12.67& 295.00& 89.82& 20.45& 15.95& 209.05 \\
           & \#945 & 107.34 & 93.26 & 14.08 & 12.63& 11.59& 273.00& 78.06& 11.07& 9.00& 184.78 \\
           & \#097 & 108.18 & 95.20 & 12.97 & 11.33& 10.45& 274.00& 73.36& 9.80& 8.45& 185.69 \\
           & \#907 & 55.15 & 47.67 & 7.48 & 6.35& 5.81& 219.00& 50.61& 4.32& 3.77& 160.98 \\
    \hline
    \hline
8742 & \#234 & 296.59 & 260.31 & 36.29 & 29.86& 28.80& 383.00& 101.79& 15.71& 14.62& 258.68 \\
           & \#283 & 160.41 & 137.78 & 22.63 & 16.60& 16.10& 312.00& 87.69& 9.47& 8.90& 235.18 \\
           & \#254 & 147.26 & 128.57 & 18.69 & 16.98& 15.37& 303.00& 100.62& 17.43& 12.99& 190.07 \\
           & \#092 & 75.03 & 64.57 & 10.46 & 9.14& 8.46& 242.00& 84.18& 9.84& 8.52& 151.06 \\
    \hline
    \hline
8743 & \#238 & 327.12 & 289.42 & 37.70 & 33.03& 30.92& 396.00& 116.33& 24.41& 20.78& 255.62 \\
           & \#328 & 66.83 & 56.77 & 10.06 & 8.63& 8.36& 233.00& 56.61& 5.64& 5.44& 167.70 \\
           & \#515 & 56.33 & 48.75 & 7.58 & 6.79& 6.40& 220.00& 56.20& 4.66& 4.34& 158.78 \\
           & \#437 & 52.65 & 45.23 & 7.42 & 6.78& 6.31& 215.00& 54.73& 8.00& 6.83& 152.63 \\
           & \#421 & 36.75 & 31.44 & 5.31 & 4.69& 4.29& 191.00& 46.77& 4.33& 3.82& 141.13 \\
    \hline
    \hline
    \hline
8716 & \#173 & 753.34 & 673.07 & 80.28 & 64.75& 54.89& 523.00& 230.70& 39.85& 24.19& 300.90 \\
           & \#253 & 312.64 & 281.54 & 31.10 & 27.42& 24.03& 390.00& 109.32& 8.85& 6.72& 261.01 \\
           & \#581 & 170.89 & 151.97 & 18.92 & 17.26& 14.68& 319.00& 100.06& 8.65& 6.07& 207.37 \\
           & \#296 & 153.65 & 135.26 & 18.39 & 16.36& 13.34& 308.00& 93.98& 8.39& 5.07& 205.79 \\
    \hline
    \hline
8717 & \#317 & 739.03 & 673.61 & 65.41 & 62.44& 56.86& 519.00& 157.32& 30.96& 24.15& 345.96 \\
           & \#288 & 280.20 & 249.36 & 30.85 & 28.17& 24.27& 376.00& 108.00& 14.90& 9.54& 245.54 \\
           & \#348 & 222.99 & 197.69 & 25.30 & 23.15& 19.23& 348.00& 111.59& 14.26& 8.81& 222.23 \\
    \hline
    \hline
8721 & \#011 & 271.20 & 237.67 & 33.53 & 28.36& 26.61& 372.00& 105.34& 9.83& 7.90& 248.27 \\
           & \#945 & 107.97 & 94.22 & 13.75 & 12.44& 10.20& 274.00& 78.03& 5.60& 3.73& 187.11 \\
           & \#097 & 106.32 & 93.99 & 12.33 & 10.54& 8.99& 272.00& 73.94& 5.14& 3.83& 185.74 \\
    \hline
    \hline
8722 & \#293 & 729.97 & 658.54 & 71.42 & 65.54& 58.46& 516.00& 151.34& 28.27& 20.95& 332.75 \\
           & \#234 & 292.30 & 259.61 & 32.69 & 27.44& 25.72& 381.00& 100.25& 7.96& 6.87& 261.70 \\
           & \#283 & 157.20 & 136.37 & 20.83 & 15.78& 13.95& 310.00& 86.88& 4.84& 3.62& 237.45 \\
           & \#254 & 145.18 & 127.78 & 17.40 & 16.04& 13.13& 302.00& 102.94& 6.47& 4.33& 193.14 \\
    \hline
    \hline
8723 & \#647 & 773.24 & 689.69 & 83.55 & 71.59& 59.77& 527.00& 182.08& 92.82& 44.22& 325.39 \\
           & \#238 & 318.02 & 285.59 & 32.44 & 29.60& 24.81& 392.00& 107.38& 12.58& 8.06& 260.00 \\
           & \#563 & 271.83 & 240.05 & 31.78 & 27.93& 24.49& 372.00& 110.79& 24.03& 15.38& 240.78 \\
    \hline
    \hline
 
  \end{tabular}
\end{center}
 
\medskip
Masses are given in 10$^{10}$~M$_\odot$, distances in kpc, velocity dispersion in km~s$^{-1}$.
\end{table*}

\begin{figure}
  \begin{center}
    \includegraphics[width=.45\textwidth]{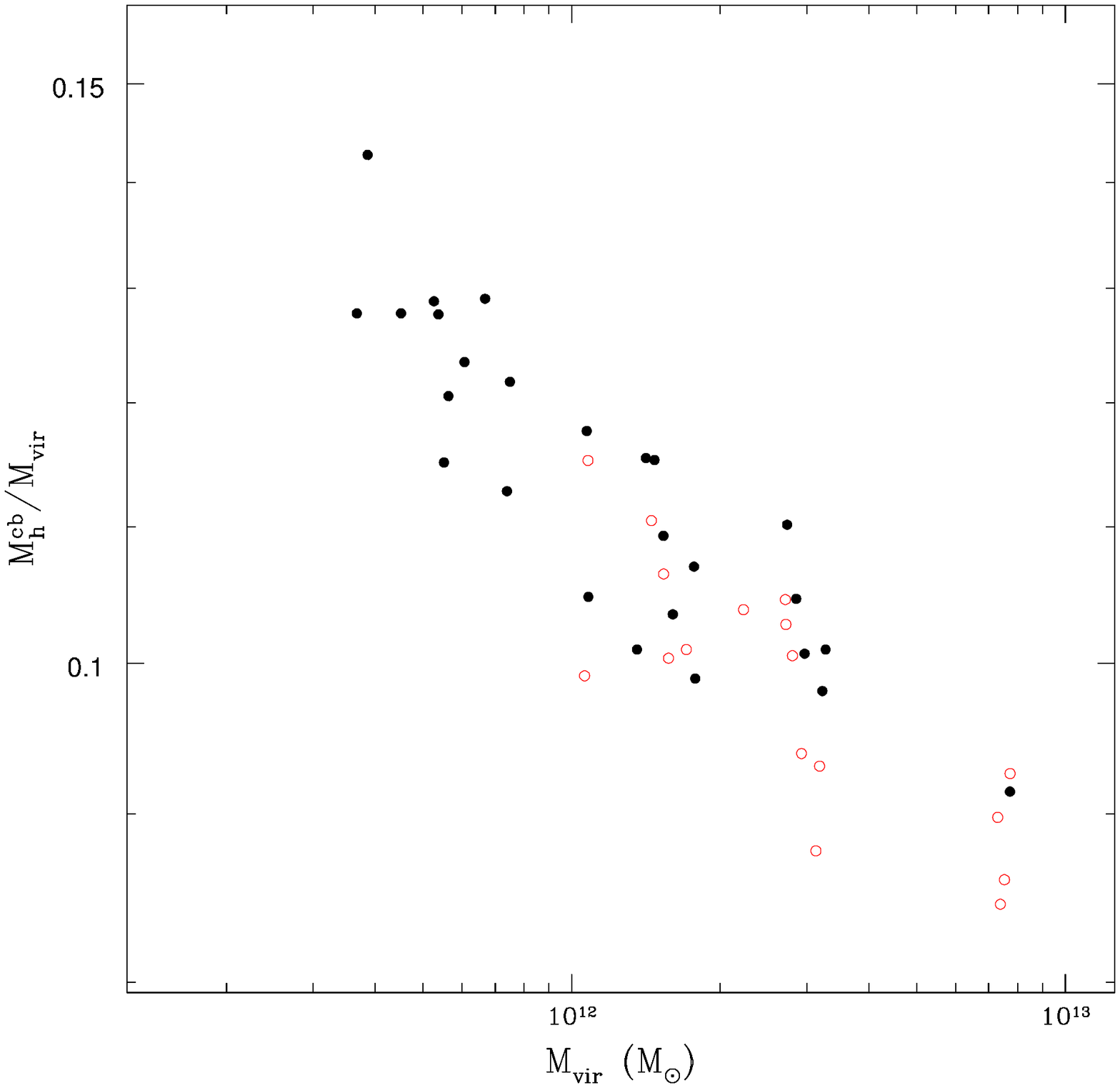}
    \caption{
      Masses of cold baryons  inside the virial
      radii  in units 
      of the corresponding virial mass for the ELO sample.
    filled black symbols: SF-A sample ELOs; open red symbols: SF-B sample ELOs.
 }
    \label{CocMcoldMvir}
  \end{center}
\end{figure}

\begin{figure}
  \begin{center}
      \includegraphics[width=.45\textwidth]{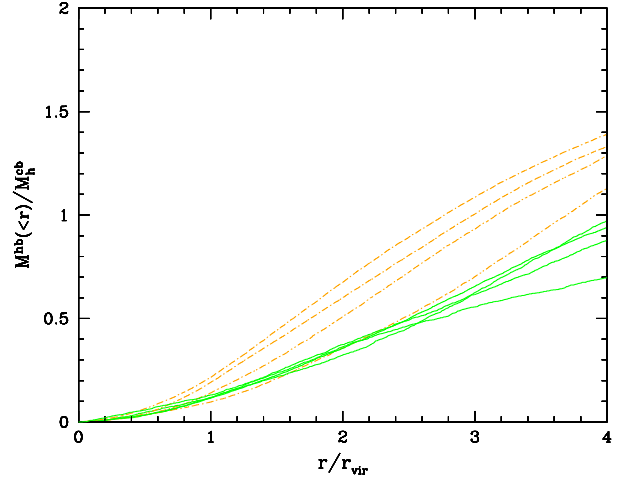}
      \caption{ The hot gas density profile normalised to the mass of 
      cold baryons inside the virial radii for isolated ELOs. Orange
      point-dashed lines: ELOs
  in the mass interval $1.5 \times 10^{12}$ M$_{\odot} \leq M_{\rm vir} < 5 \times 10^{12}$ M$_{\odot}$;
 continuous green lines: ELOs in the mass interval $M_{\rm vir} < 1.5 \times 10^{12}$ M$_{\odot}$.
}      
\label{HgasMcbV}
\end{center}
\end{figure}

An important point is the  amount
of  gas infall \emph{relative} to the halo mass scale.
As  illustrated in Figure~\ref{CocMcoldMvir}
for $M_{\rm h}^{\rm cb}/ M_{\rm vir}$, any of the ratios
$M_{\rm h}^{\rm ab}/ M_{\rm vir}$,
$M_{\rm h}^{\rm cb}/ M_{\rm vir}$ or $M_{\rm h}^{\rm star}/ M_{\rm vir}$
decreases
as $M_{\rm vir}$ increases, as observationally found at smaller scales
(see $\S$\ref{Intro}). Note that we have  in any case
$M_{\rm h}^{\rm ab}/ M_{\rm vir} < \Omega_b / \Omega_m = 0.171$,
the average cosmic fraction, so that there is a lack of baryons
 within $r_{\rm vir}$ relative to the
dark mass content that becomes more important as  $ M_{\rm vir}$ increases.
 Otherwise,
 heating  processes along ELO assembly  give rise to a hot gas
halo around the objects, partially beyond
the virial radii.  The amount of hot gas mass outside the virial radii,
normalised to the ELO stellar mass $M_{\rm bo}^{\rm star}$
(see $\S$\ref{ResBarObj}), increases with the mass scale.
It also increases relative to
the cold gas content at the halo scale.
To illustrate this point, in Figure~\ref{HgasMcbV} we plot the integrated
hot gas density profile normalised to the mass of cold baryons inside 
the virial radii, $M_{\rm h}^{\rm cb}$. The mass effect can be 
clearly appreciated in this Figure, where we
see the following:

(i) The mass of hot gas increases
monotonically up to $r \simeq 4 r_{\rm vir}$, and maybe also beyond this value,
but it is difficult at these large radii to properly dilucidate
whether or not a given hot gas mass element belongs to a given ELO or to another
close one (to alleviate this difficulty, only those ELOs not having
massive neighbours within  radii of  6$\times r_{\rm vir}$
have been considered to draw this Figure).
 
(ii) The hot gas mass fraction increases
with $M_{\rm vir}$ at given $r/r_{\rm vir}$.
This suggests that the cold baryons that massive ELOs miss inside $r_{\rm vir}$
relative to less massive ones appear as a diffuse warm component
at the outskirts of their configurations.

We now comment on length scales.
The overall half-mass radii $r_{\rm e, h}^{\rm tot}$
are closely correlated to $M_{\rm vir}$.
 Concerning baryon
 mass distributions, 
dissipation in shocks and gas cooling play now
important roles to determine these  mass distributions.
And so,  
the $ r_{\rm e, h}^{\rm star}$  radii
depend on how much energy  was radiated    before gaseous
particles
became  dense enough  to be turned into stars.
This, in turn, depends
on the mass scale, on the one hand, and, in a given mass range,
on the values of SF parameters, on the other hand.
And so, more massive ELOs tend to have larger $ r_{\rm e, h}^{\rm star}$  radii
and, in a given mass range, SF-A sample ELOs
tend to have larger  $ r_{\rm e, h}^{\rm star}$  radii
than their SF-B sample counterparts,
because the SF implementation in the code demands  denser gas
to form stars in the later than in the former.
This effect is more remarkable for sizes at the scale of the
baryonic object, as we shall see in the next subsection.

\subsection{Masses and Sizes at the Scale of the Baryonic Object}
\label{ResBarObj}
 
Let us now turn to the study of ELOs at the scale of the
baryonic objects themselves, that is, at scales of some tens of kpcs.
Physically, the   mass parameter at the ELO  scale
is $M_{\rm bo}^{\rm cb}$,
the total amount of cold baryons that have reached
the central   volume of the haloes,
forming an ELO.
Most of these cold baryons
have turned into stars, depending on
the strength of the dynamical  activity in the
volume surrounding the  proto-ELO
at high $z$, and, also, on the values of the SF parameters.
$M_{\rm bo}^{\rm star}$ is the stellar mass.
It can be estimated from
 luminosity data  through modelling (see Kauffmann et al.
  2003a, for example\footnote{The results
  of these authors indicate that for SDSS
     elliptical galaxies the
      stellar-mass-to-light ratio, $\gamma^{\rm star}$,
      can be taken to be constant in the
       range of absolute luminosities $M < -21$
        (see Kauffmann et al. 2003b).
	 The values of the logarithm
	  of this ratio are 
	  $\log \gamma^{\rm star}_{\it r} \simeq 0.53$ and 
	  $\log \gamma^{\rm star}_{\it z} \simeq 0.25$,
	   with dispersions $\sigma_S < 0.15$ and 0.1,  in the
$\textit{r}$ and $\textit{z}$ SDSS bands, respectively.
}).
    Effective or half-mass radii
      at the baryonic object scale,
       $r_{\rm e, bo}^{\rm cb}$ and
         $r_{\rm e, bo}^{\rm star}$, can be
	  defined as those radii enclosing half the $M_{\rm bo}^{\rm cb}$
	   or $M_{\rm bo}^{\rm star}$ masses, respectively.
	    These are the relevant  size scales for the
	     intrinsic ELOs, but the observationally relevant size scales
	      are the {\it projected} half-mass radii.
	       They are
	        determined  from $M_{\rm cyl}(R)$, the integrated projected mass
		 density in concentric cylinders of radius $R$ for the different
		 constituents. For example,
		   $R_{\rm e, bo}^{\rm cb}$ and
		    $R_{\rm e, bo}^{\rm star}$ are the projected radii
		     where
		      $M^{\rm  cb}_{\rm cyl}(R)$ and
		        $M^{\rm star}_{\rm cyl}(R)$
are equal to $M_{\rm bo}^{\rm cb}/2$
  and  to  $M_{\rm bo}^{\rm star}/2$, respectively. 
Significant parameters at the baryonic object scale
  are listed in Table~\ref{tabetloscale}.
Note that ELOs have a  lower limit
 in their stellar mass content of 3.8 $\times$ 10$^{10}$ M$_{\odot}$
 (see Kauffmann et al. 2003b for a similar result in SDSS early-type galaxies).

\begin{table}

  \caption[Masses and sizes of ELOs at the object scale]
  {Masses, sizes and mean square velocities at the baryonic object scale,
  as well as projected radii and central stellar l.o.s.
  velocity dispersions ($z=0$)}
  \label{tabetloscale}
 
  \begin{center}
    \footnotesize
 
  \begin{tabular}{lccccccc}
    \hline
  Run&  ELO&$M_{\rm bo}^{\rm cb}$&$M_{\rm bo}^{\rm star}$&$r_{\rm e, bo}^{\rm star}$&$R_{\rm e, bo}^{\rm star}$&$\sigma_{\rm 3, bo}^{\rm star}$&$\sigma_{\rm los, 0}^{\rm star}$\\
    \hline
    \hline
8714 & \#173 & 43.12 & 42.59 & 13.01 & 12.72& 351.93 &226.67 \\
           & \#353 & 25.62 & 25.17 & 8.25 & 7.95& 297.11 &192.47 \\
           & \#581 & 13.72 & 13.45 & 6.75 & 6.66& 222.42 &136.18 \\
           & \#296 & 12.21 & 11.95 & 4.76 & 4.69& 225.95 &138.39 \\
           & \#373 & 7.38 & 7.23 & 3.85 & 3.83& 186.95 &116.88 \\
           & \#772 & 5.62 & 5.51 & 2.33 & 2.30& 181.59 &107.28 \\
           & \#284 & 5.70 & 5.48 & 2.81 & 2.74& 169.09 &103.74 \\
    \hline
    \hline
8747 & \#288 & 20.11 & 20.05 & 7.84 & 7.81& 269.58 &171.50 \\
           & \#115 & 12.21 & 12.17 & 2.94 & 2.96& 259.05 &161.63 \\
           & \#189 & 4.74 & 4.68 & 2.57 & 2.61& 169.93 &104.81 \\
           & \#915 & 4.52 & 4.32 & 2.23 & 2.19& 164.69 &99.61 \\
    \hline
    \hline
8741 & \#011 & 19.10 & 19.10 & 5.36 & 5.35& 279.61 &174.62 \\
           & \#017 & 10.34 & 10.15 & 4.15 & 4.09& 224.37 &139.54 \\
           & \#930 & 8.95 & 8.85 & 4.68 & 4.70& 214.97 &138.64 \\
           & \#945 & 10.00 & 9.89 & 4.58 & 4.62& 209.68 &128.84 \\
           & \#097 & 9.46 & 9.34 & 5.00 & 5.05& 206.21 &132.19 \\
           & \#907 & 5.63 & 5.55 & 2.57 & 2.60& 187.67 &119.71 \\
    \hline
    \hline
8742 & \#234 & 27.42 & 27.13 & 9.33 & 9.23& 287.18 &190.47 \\
           & \#283 & 13.51 & 13.44 & 4.28 & 4.30& 246.54 &151.56 \\
           & \#254 & 10.91 & 10.81 & 4.50 & 4.50& 221.72 &134.23 \\
           & \#092 & 6.55 & 6.38 & 3.14 & 3.17& 177.69 &108.46 \\
    \hline
    \hline
8743 & \#238 & 20.97 & 20.88 & 6.76 & 6.79& 289.43 &181.53 \\
           & \#328 & 7.76 & 7.60 & 3.40 & 3.36& 187.15 &116.49 \\
           & \#515 & 5.74 & 5.63 & 2.60 & 2.62& 185.04 &114.13 \\
           & \#437 & 4.79 & 4.70 & 2.36 & 2.40& 172.71 &105.13 \\
           & \#421 & 3.91 & 3.81 & 2.11 & 2.14& 161.41 &102.09 \\
    \hline
    \hline
8716 & \#173 & 39.13 & 38.07 & 7.07 & 6.75& 355.08 &237.07 \\
           & \#253 & 22.98 & 22.54 & 4.40 & 4.28& 305.37 &199.34 \\
           & \#581 & 13.03 & 12.46 & 3.35 & 3.18& 243.71 &152.89 \\
           & \#296 & 10.87 & 10.43 & 2.52 & 2.50& 241.58 &140.14 \\
    \hline
    \hline
8717 & \#317 & 42.28 & 41.55 & 8.12 & 7.77& 387.30 &223.73 \\
           & \#288 & 18.63 & 18.25 & 3.48 & 3.45& 286.59 &169.81 \\
           & \#348 & 15.36 & 14.84 & 3.73 & 3.57& 261.06 &158.31 \\
    \hline
    \hline
8721 & \#011 & 21.73 & 21.37 & 3.34 & 3.31& 287.47 &177.04 \\
           & \#945 & 9.99 & 9.30 & 2.55 & 2.44& 226.33 &128.90 \\
           & \#097 & 8.62 & 8.34 & 2.34 & 2.36& 213.00 &133.46 \\
    \hline
    \hline
8722 & \#293 & 43.56 & 42.75 & 9.40 & 9.09& 365.95 &215.68 \\
           & \#234 & 22.40 & 22.28 & 3.53 & 3.57& 313.03 &196.62 \\
           & \#283 & 12.67 & 12.51 & 2.10 & 2.13& 262.81 &163.89 \\
           & \#254 & 11.97 & 11.52 & 2.64 & 2.62& 237.10 &137.68 \\
    \hline
    \hline
8723 & \#647 & 34.01 & 33.43 & 5.86 & 5.76& 336.69 &210.25 \\
           & \#238 & 20.09 & 19.91 & 3.78 & 3.74& 297.51 &191.05 \\
           & \#563 & 13.67 & 13.44 & 2.41 & 2.45& 274.84 &164.01 \\
    \hline
    \hline
 
  \end{tabular}
\end{center}
 
\medskip
Masses are given in 10$^{10}$~M$_\odot$, distances in kpc,
velocity dispersions in km~s$^{-1}$.
\end{table}

\begin{figure}
  \begin{center}
    \includegraphics[width=.45\textwidth]{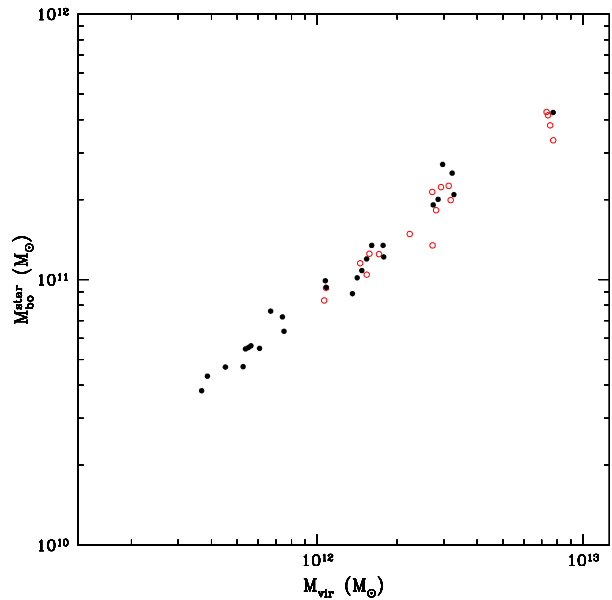}
    \caption[Cold baryon and stellar masses at the baryonic object scale]{
      Stellar  masses at the baryonic object scale
       scale versus halo mass for the ELO sample. Symbols are as in previous
      Figures
      }
    \label{CorrMobMvir}
  \end{center}
\end{figure}

\begin{figure}
  \begin{center}
    \includegraphics[width=.45\textwidth]{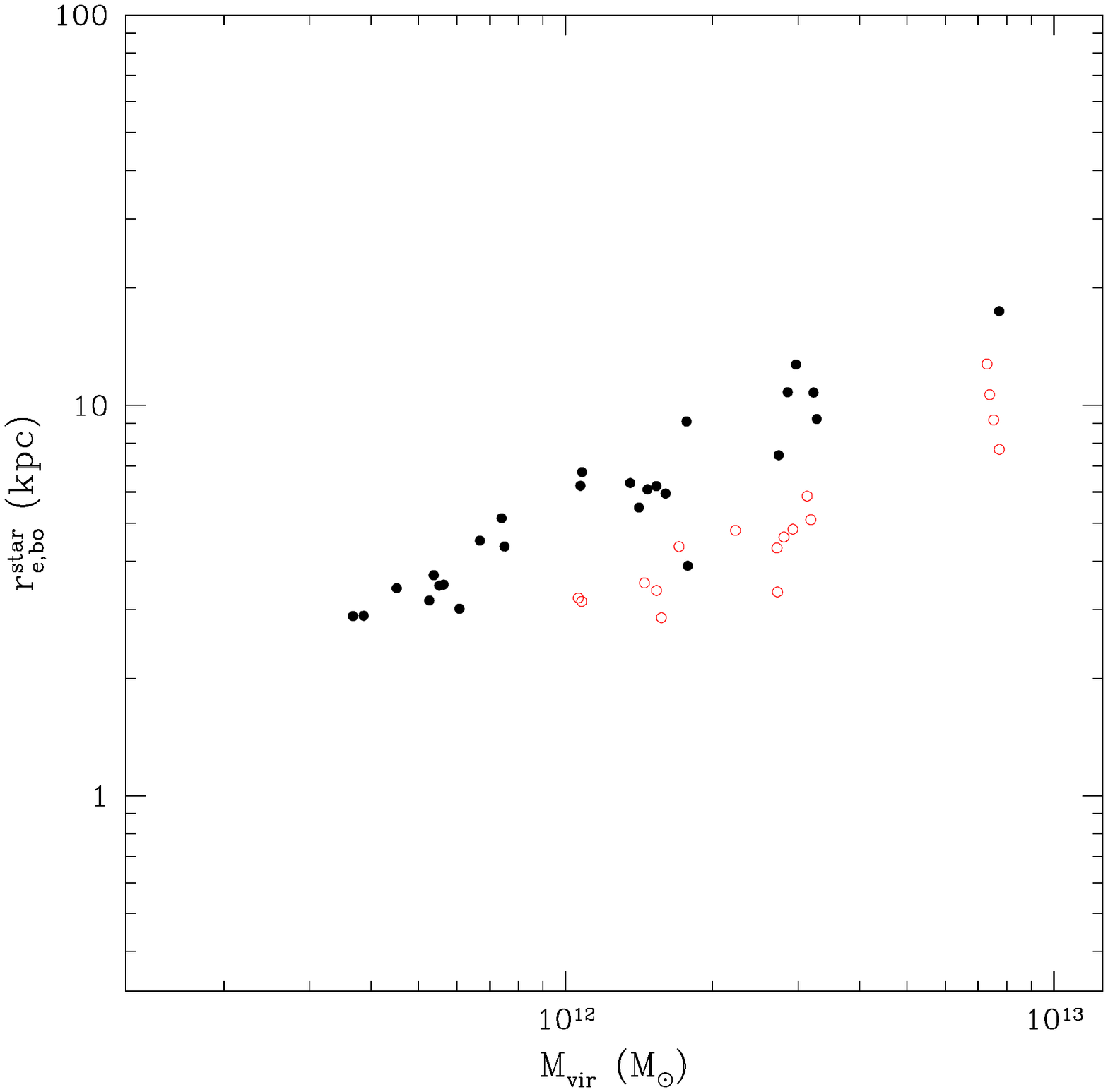}
    \caption[Half-mass radii for  stellar masses versus halo mass]{
      The 3D half-mass radii for   stellar 
      masses at the baryonic object scale versus halo mass for the ELO sample.
      Symbols are as in previous Figures
      }
    \label{CorrRobMvir}
  \end{center}
\end{figure}

To illustrate  how the halo total mass,
$M_{\rm vir}$, determines the ELO structure at kpc scales,
in Figures~\ref{CorrMobMvir}
and~\ref{CorrRobMvir}  we draw 
$M_{\rm bo}^{\rm star}$  and
$r_{\rm e, bo}^{\rm star}$ versus
$M_{\rm vir}$, respectively, for the ELO sample.
A good correlation is  apparent in Figure~\ref{CorrMobMvir},
where it is shown that ELO  stellar masses are mainly
determined by the halo mass scale, $M_{\rm vir}$, with only
a very slight  dependence on the SF parametrisation
(SF-A type ELOs have a slightly higher stellar content
than their SF-B counterparts, as expected).             
Figure~\ref{CorrRobMvir}  shows also  a good correlation
between the length scales for the  stellar masses
and $M_{\rm vir}$, but now the  sizes  depend also on
 the SF parameters.       
The physical foundations of this behaviour  are the same
as discussed in $\S$\ref{MaSizes}.

\begin{figure}
  \begin{center}
      \includegraphics[width=.45\textwidth]{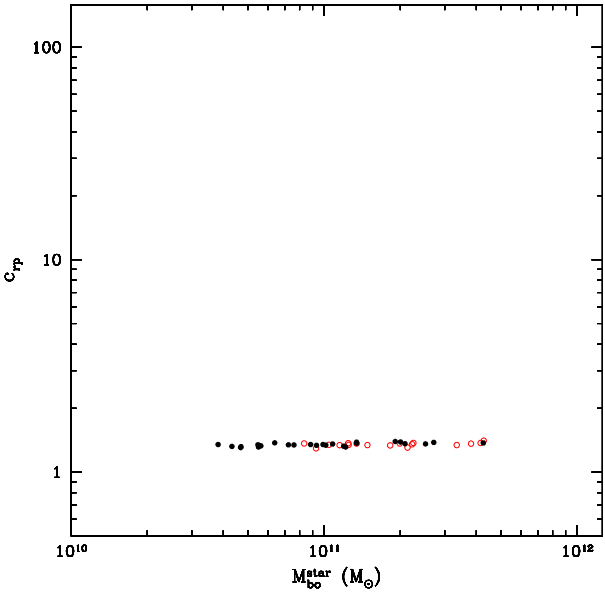}
        \caption{The $c_{\rm rp} \equiv r_{\rm e, bo}^{\rm star} / R_{\rm e, bo}^{\rm star}$ ratios versus the stellar
	      masses at the baryonic object scale. Symbols are as in previous Figures.
}
   \label{Crp4.11}
 \end{center}
\end{figure}

The observationally relevant scalelengths are the 
projected half-mass radii
 $R_{\rm e, bo}^{\rm star}$.
  Their correlations with 
 their intrinsic  three dimensional  counterparts
  $r_{\rm e, bo}^{\rm star}$ are very good, as illustrated
in Figure~\ref{Crp4.11}, 
where the very low dispersion in the plots of the
$c_{\rm rp} \equiv r_{\rm e, bo}^{\rm star} / R_{\rm e, bo}^{\rm star}$
ratios   versus the stellar mass $M_{\rm bo}^{\rm star}$ can be appreciated.
The results of a fit to a power law of the form 
$c_{\rm rp} = A_{\rm rp} (M_{\rm bo}^{\rm star})^{\beta_{\rm rp}}$
\footnote{We list the different ratio definitions we use in this paper and their corresponding
logarithmic slopes in Table~\ref{tabnombres2}}
are given in Table \ref{TabFP}
, where we see that the $c_{\rm rp} $
ratios show a very mild mass dependence in the SF-A sample and none in the
SF-B sample.
This result is important because it indicates
that the observationally available {\it projected} radii 
$R_{\rm e, bo}^{\rm star}$ are
robust estimators of the physically meaningful size scales
$r_{\rm e, bo}^{\rm star}$.

\begin{figure}
  \begin{center}
    \includegraphics[width=.45\textwidth]{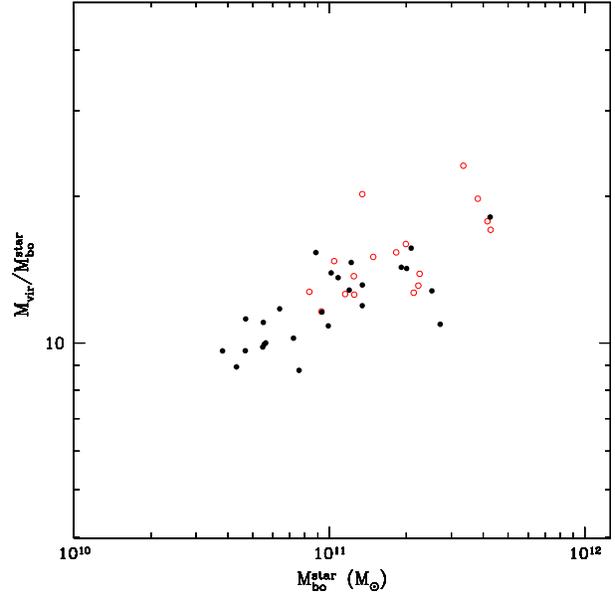}
    \caption{
    The $M_{\rm vir}/M_{\rm bo}^{\rm star}$ ratios as a function of
                          the ELO mass scale. Symbols are as in previous Figures
}
    \label{CMvirMob4.5}
  \end{center}
\end{figure}

We now address the correlations of normalised mass and size scales. 
The increasing behaviour  of the $M_{\rm vir}/ M_{\rm bo}^{\rm cb}$
and $M_{\rm vir}/ M_{\rm bo}^{\rm star}$ ratios with increasing
mass scale are very  interesting. In particular, 
the last ratio (Figure~\ref{CMvirMob4.5}) follows the same
trends as the empirical $M_{\rm e}/L$ versus $L$ relation, see Bernardi et al. 2003b.
The results of a fit to a power law of the form
$ M_{\rm vir}/M_{\rm bo}^{\rm star} = A_{\rm vir} (M_{\rm bo}^{\rm star})^{\beta_{\rm vir}}$ 
 are given in Table \ref{TabFP},
 where we see that they do not
 depend on the SF parameterisation.

\begin{table}
\caption[Slopes for Linear Fits]{Slopes for Linear Fits}
\label{TabFP}
\begin{center}
\begin{tabular}{lcccc}
\hline
&SF-A &&SF-B  &\\
\hline
\hline
$M_1$               & 0.256 & $\pm$ 0.035& 0.281 & $\pm$ 0.048      \\
$\beta_{\rm vir}$ & 0.221 &   $\pm$ 0.083 & 0.237 & $\pm$ 0.158    \\
$\beta_{\rm M }$  &-0.204 & $\pm$ 0.116 & -0.247& $\pm$ 0.189 \\
\hline
\hline
$\beta_{\rm F}$   & 0.025 & $\pm$ 0.048 & 0.022 & $\pm$ 0.081 \\
$\beta_{\rm vd}$  & 0.021 & $\pm$ 0.041 & 0.076 & $\pm$ 0.075 \\
$\beta_{\rm vpc}$ &-0.044 & $\pm$ 0.029 &-0.044 & $\pm$ 0.093 \\
$\beta_{\rm rd}$  &-0.225 & $\pm$ 0.127 &-0.316 & $\pm$ 0.199 \\
$\beta_{\rm rp}$  & 0.019 & $\pm$ 0.009 & 0.016 & $\pm$ 0.017 \\
\hline
\hline

\end{tabular}
\end{center}

\medskip

Column 2: the slopes of the $\kappa_{3}^{\rm D} = M_1 \kappa_{1}^{\rm D} + M_0$
relation (direct fits);
the slopes of the $M_{\rm vir}/M_{\rm bo}^{\rm star}$
and
$c_{\rm i} \propto (M_{\rm bo}^{\rm star})^{\beta_{\rm i}}$
scaling relations for the the SF-A sample,
calculated in $\log - \log$ plots through direct fits.
Column 3: their respective
95\% confidence intervals. Columns 4 and 5: same as columns 2 and 3
for the SF-B sample.

\end{table}

\begin{figure}
  \begin{center}
      \includegraphics[width=.45\textwidth]{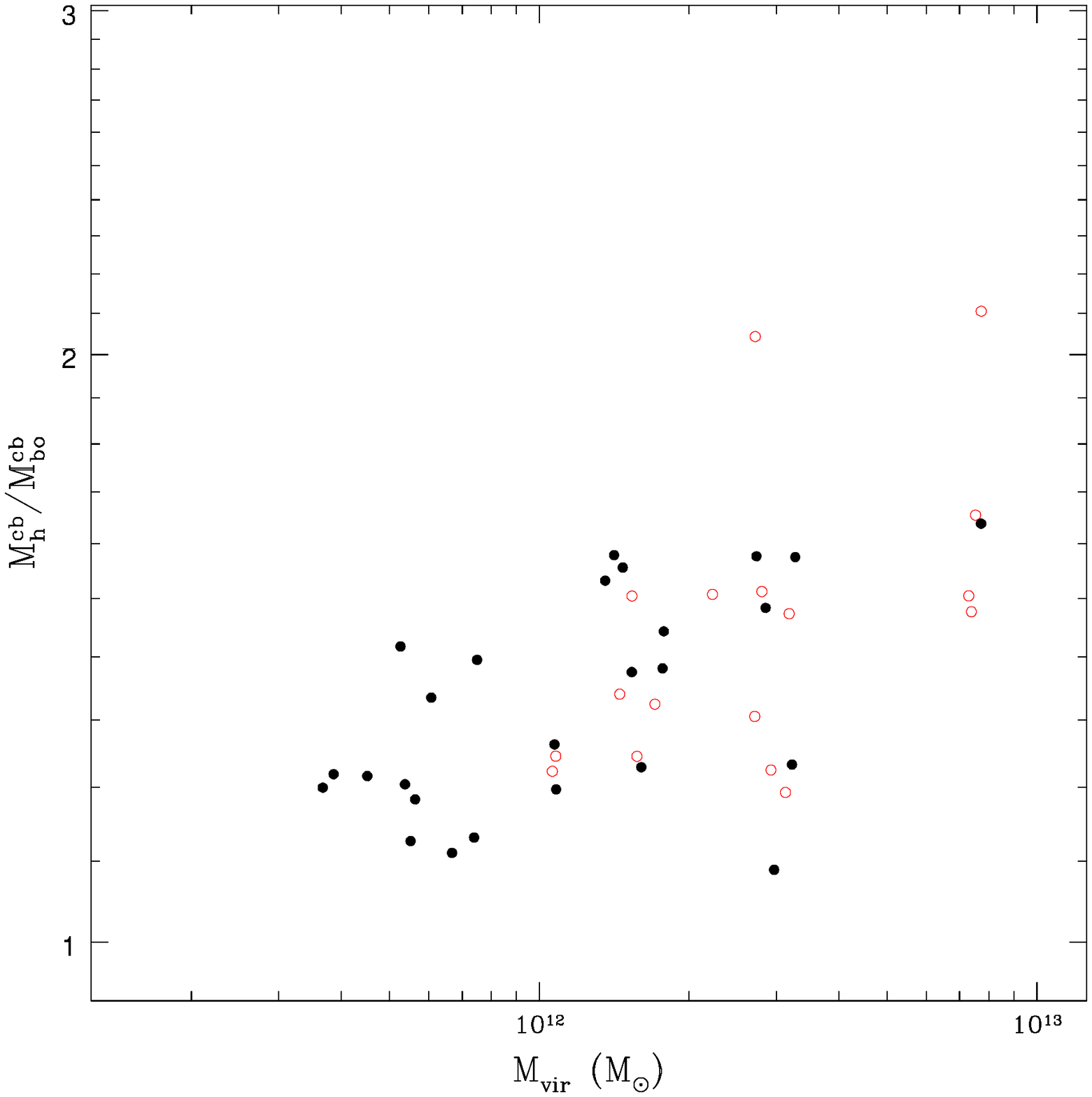}
      \caption{
      The $M_{\rm h}^{\rm cb}/M_{\rm bo}^{\rm cb}$ ratios as a function of
      the ELO mass scale. Symbols are as in previous Figures
      }
      \label{CocMcbhbo}
    \end{center}
\end{figure}

To have an idea on how important cold baryon infall has been at the baryonic object scale
relative to that at the halo scale, in Figure~\ref{CocMcbhbo}
the $M_{\rm h}^{\rm cb}/M_{\rm bo}^{\rm cb}$ ratios are drawn as a function of
      the ELO mass scale. We see that in any case more than half the mass of
      cold baryons inside the virial radii are concentrated in the central
      baryonic object, and that there is a  mass effect in the sense
      that this fraction grows with decreasing ELO mass scale,
      and no appreciable SF effect.

\begin{figure}
  \begin{center}
      \includegraphics[width=.45\textwidth]{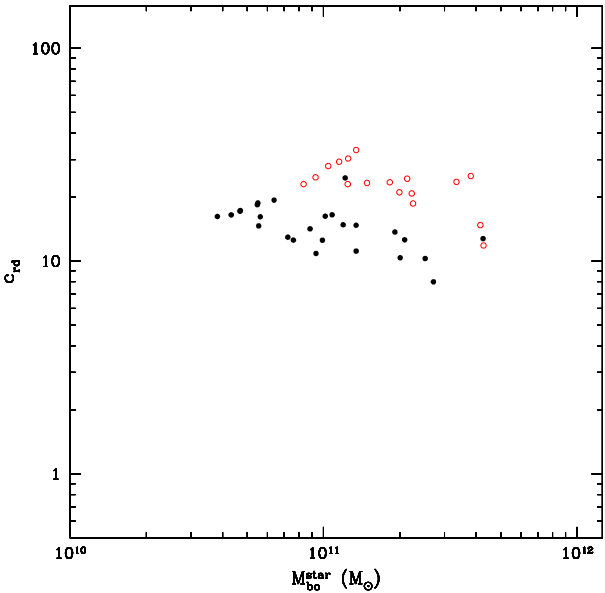}
          \caption{
 The $c_{\rm rd} \equiv r_{\rm e, h}^{\rm tot} / r_{\rm e, bo}^{\rm star}$
  ratios as a function of
        the ELO mass scale. Symbols are as in previous Figures.
	      Spatial homology breaking is clear in this Figure
}
  \label{crd4.10}
 \end{center}
\end{figure}

Concerning sizes,
in Figure~~\ref{crd4.10} we plot the 
$c_{\rm rd} \equiv r_{\rm e, h}^{\rm tot} / r_{\rm e, bo}^{\rm star}$
ratios versus the  $M_{\rm bo}^{\rm star}$ mass scale
for ELOs in both the SF-A  and the SF-B samples.
In this Figure the effects of SF parameterisation are clear:
SF-A type ELOs have larger sizes relative to the halo size
than SF-B type ELOs.
There is also a clear mass effect, with more massive ELOs
less concentrated relative to the total mass distribution
than less massive ones
(i.e., spatial homology breaking; note, however that the
scatter is important). Moreover, Figure~~\ref{crd4.10}
suggests that this trend does not significantly depend
on the SF parametrisation. These indications are quantitatively
confirmed through a fit to a power law
$c_{\rm rd} = A_{\rm rd} (M_{\rm bo}^{\rm star})^{\beta_{\rm rd}}$
(see Table \ref{TabFP}) and have interesting implications to explain the
tilt of the observed FP.

\section{Kinematics}
\label{Kine}

\subsection{Three-dimensional velocity distributions}
\label{3DVelDis}

\begin{figure}
\begin{center}
\includegraphics[width=.45\textwidth]{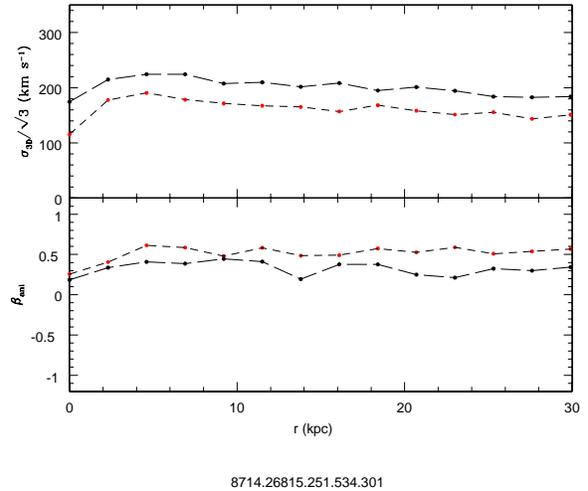}
\caption{
The $\sigma_{\rm 3D}(r)$ profiles
of a typical ELO in the SF-A sample.
Also shown are their anisotropy profiles $\beta_{\rm ani}(r)$.
Long-dashed lines: dark matter;
short-dashed lines: stars
}
\label{Sig3DAni}
\end{center}
\end{figure}

Shapes and mass density profiles (i.e., positions)
are  related to the 3D velocity distributions
of relaxed  E  galaxies through
 the Jeans equation (see Binney \& Tremaine 1987).
 Observationally, the informations
 on such 3D distributions is not available for external galaxies,
 only the line-of-sight velocity distributions
 (LOSVD) can be inferred from their spectra. They have been found to be
 close to gaussian (Binney \& Tremaine 1987; van der Marel \& Franx 1993),
 so that simple equilibrium models can be expected to
 adequately describe their dynamical state (de Zeeuw \& Franx 1991).
 The complete six dimensional phase space informations
 for each of the particles sampling the ELOs provided
 by numerical simulations,
 allow us  to calculate the
 velocity profiles, $V_{\rm cir}(r)$, the 3D
 profiles for the velocity dispersion, $\sigma_{\rm 3D}(r)$,
and their corresponding anisotropy profiles. The anisotropy is 
defined as:

\begin{equation}
\beta_{\rm ani} = 1 - \frac{\sigma^2_{\rm t}}{2\,\sigma^2_{\rm r}},
\label{AniDef}
\end{equation}

where $\sigma_{\rm r}$ and $\sigma_{\rm t}$ are the radial and
tangential velocity dispersions ($\sigma_{\rm t}^2 = \sigma_{\theta}^2 + \sigma_{\phi}^2$), relative to the centre of the
object.
These profiles, as well as the LOS velocity
$V_{\rm los}(R)$ and LOS velocity dispersion $\sigma_{\rm los}(R)$
profiles,
are analysed in detail in O\~norbe et al., in preparation.
Figure~\ref{Sig3DAni}
illustrates the more outstanding results we have obtained:
ELO velocity dispersion profiles in three dimensions
are slightly decreasing for increasing $r$, both for
dark matter and stellar particles, $\sigma^{\rm dark}_{3D}(r)$
and $\sigma^{\rm star}_{3D}(r)$.
It has been found that
$(\sigma^{\rm dark}_{3D}(r))^2 \sim$ (1.4 -- 2)
$(\sigma^{\rm star}_{3D}(r))^2$,
as Loewenstein 2000 had found on theoretical grounds.
This is so because stars are formed from gas that had lost
energy by cooling.
This result on kinematical segregation is very interesting
because it has the implication that the use of stellar
kinematics to measure the total mass of ellipticals
could result into inaccurate values.
The study of anisotropy
has shown that it is always positive and almost
non-varying with $r$ (recall, however, that ELOs
are non-rotating and that they have been identified as
dynamically relaxed objects, so that there are not recent mergers
in our samples). The stellar component generally shows
more anisotropy than
the dark component, maybe coming from the radial motion of the gas particles
that gave rise to the stars.

\subsection{Global  parameters for the velocity distribution}
\label{GloLOSVDPar}
  
\begin{figure}
  \begin{center}
      \includegraphics[width=.45\textwidth]{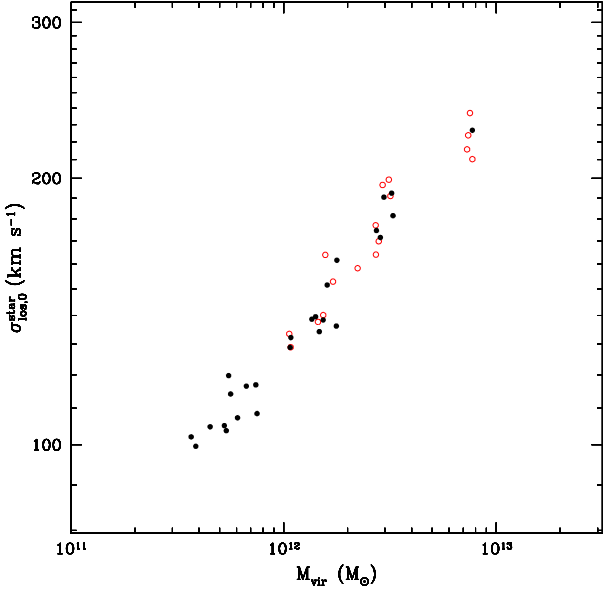}
	\caption[L.o.s.\  velocity dispersion versus virial mass]{
 The correlation between the central l.o.s.\  velocity dispersion
			           and the virial mass for the ELO samples.
				   Symbols are as in previous Figures.
}
\label{CorrSlos0Svir}
\end{center}
\end{figure}

Only for a limited number of ellipticals are the
$V_{\rm los}(R)$ or $\sigma_{\rm los}(R)$
profiles available. Observationally,
a useful characterisation of the
velocity dispersion of an E galaxy is
provided by its central stellar
line-of-sight velocity dispersion, $\sigma_{\rm los, 0}$.
It corresponds to the velocity dispersion of the stellar
(as opposed to dark matter or other) component.
Due to its interest, $\sigma_{\rm los, 0}$ has deserved an
important attention in literature and it had been measured
for several E  galaxy samples before the SDSS results
(Faber et al.\ 1987;
 Djorgovski \& Davis 1987; Dressler et al.\ 1987;
Lucey, Bower \& Ellis 1991; J{\o}rgensen, Franx \& Kjoergaard 1993;
J{\o}rgensen  et al.\ 1996; Kelson et al.\ 1997;
Kelson et al.\ 2000, Bernardi et al.\ 2002).
  In Table~\ref{tabetloscale} the values of $\sigma_{\rm los, 0}^{\rm star}$
  for the ELO sample are listed\footnote{Recall that the central LOS velocity 
  dispersion for the {\it stellar} component of ELOs is
  written as $\sigma_{\rm los, 0}^{\rm star}$, with a "star"
  superindex to distinguish it 
  from that of the other ELO components}. In Figure~\ref{CorrSlos0Svir} we show the good correlation between $\sigma_{los,0}^{star}$ and the virial mass, $M_{vir}$.
  
\begin{figure}
  \begin{center}
     \includegraphics[width=.45\textwidth]{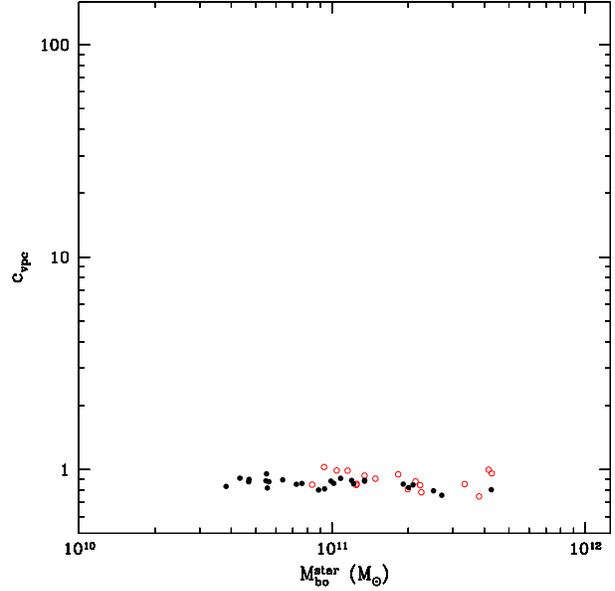}
       \caption{
$c_{\rm vpc} \equiv (\sigma_{\rm 3, bo}^{\rm star})^2 / 3 (\sigma_{\rm los, 0}^{\rm star})^2$ ratios versus the ELO mass scale.       
 Symbols are as in previous Figures
}
   \label{Cvpc4.9}
  \end{center}
\end{figure}

Physically, a measure of the average dynamical state of stars in the
ELO itself is provided by their  mean square velocity relative
to the ELO center of mass, or average three-dimensional velocity dispersion
$\sigma_{\rm 3, bo}^{\rm star}$, of which the observationally available
$\sigma_{\rm los, 0}^{\rm star}$ parameter is assumed to be a fair
estimator. To test this point, in Figure~\ref{Cvpc4.9} we plot 
the $c_{\rm vpc} \equiv (\sigma_{\rm 3, bo}^{\rm star})^2 / 3 (\sigma_{\rm los, 0}^{\rm star})^2$ ratios versus the ELO mass scale.
We see that no mass effect is apparent, and this is quantitatively
confirmed in Table~\ref{TabFP}, where the results of a fit of the
form $c_{\rm vpc} = A_{\rm vpc} (M_{\rm bo}^{\rm star})^{\beta_{\rm vpc}}$
are given. We also see that due to radial anisotropy, 
$c_{\rm vpc} < 1$, with no SF parametrisation effect.
So, there is not mass bias when using
$\sigma_{\rm los, 0}^{\rm star}$ as an estimator for 
$\sigma_{\rm 3, bo}^{\rm star}$, but some warnings are in order
concerning anisotropy effects.

\begin{figure}
\begin{center}
\includegraphics[width=.45\textwidth]{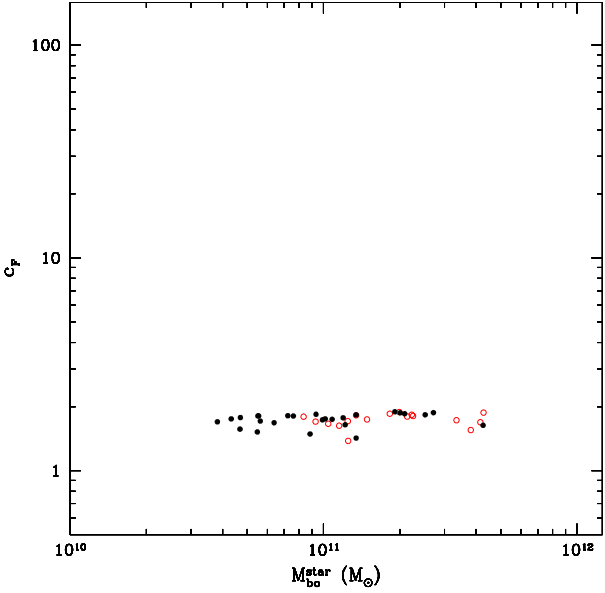}
\caption{The
$c_{\rm F} \equiv G M_{\rm vir} / (\sigma_{\rm 3, h}^{\rm tot})^2  r_{\rm e, h}^{\rm tot}$ form factors (see Eq. \ref{Virial})
versus the ELO mass scale.
Symbols are as in previous Figures. This Figure confirms that
$r_{\rm e, h}^{\rm tot}$ and $\sigma_{\rm 3, h}^{\rm tot}$
are the size and velocity dispersion ELO  parameters that must
be used in the virial theorem
}
\label{CF}
\end{center}
\end{figure}

A significant velocity dispersion parameter for ELOs at the halo scale
is $\sigma_{\rm 3, h}^{\rm tot}$, the average 3-dimensional velocity dispersion
of the whole elliptical, including both dark and baryonic matter.
According with Eq. \ref{Virial}, this is
the velocity dispersion entering in the virial theorem for the whole ELO
configuration. To test that this is in  fact the case, in Figure~\ref{CF}
we plot the  
$c_{\rm F} \equiv G M_{\rm vir} / (\sigma_{\rm 3, h}^{\rm tot})^2  r_{\rm e, h}^{\rm tot}$ form factors (see Eq. \ref{Virial}) 
as a function of $M_{\rm bo}^{\rm star}$. The lack of any significant
mass or SF parametrisation 
effects in this Figure are quantitatively confirmed through
 a fit to  power laws of the form 
$c_{\rm F} = A_{\rm F} (M_{\rm bo}^{\rm star})^{\beta_{\rm F}}$,
whose results  in Table~\ref{TabFP} are  
consistent with $c_{\rm F}$ being independent of the ELO mass scale
or SF parameter values. Note also that the $c_{\rm F}$ 
values are as expected (Binney \& Tremaine 1987). 

\begin{figure}
  \begin{center}
       \includegraphics[width=.45\textwidth]{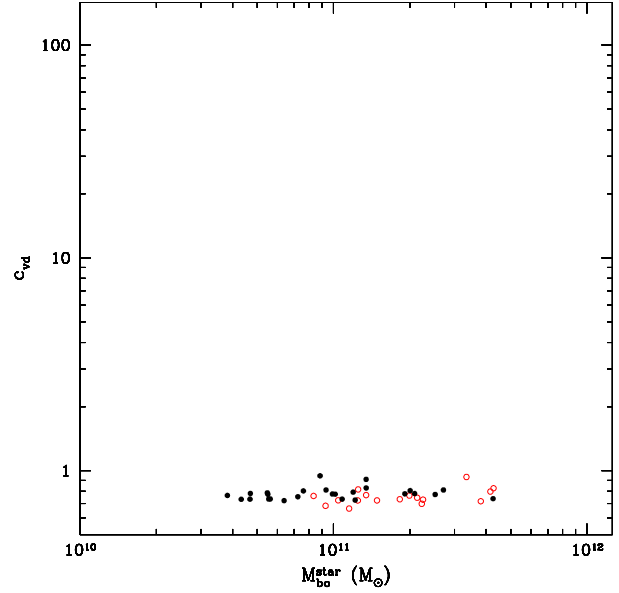}
              \caption{ The
$c_{\rm vd} \equiv (\sigma_{\rm 3, h}^{\rm tot} / \sigma_{\rm 3, bo}^{\rm star})^2$ 
ratios (average kinematical segregation)
as a function of the ELO mass scale. Symbols are as in previous Figures.
No dynamically broken homology can be seen in this Figure, but the kinematical
segregation between dark matter and stars is clear.
}
   \label{Cvd4.8}
  \end{center}
\end{figure}

Once we have confirmed that
when writting  the virial theorem for an ELO configuration,
$\sigma_{\rm 3, h}^{\rm tot}$ is the 
velocity dispersion one has to use, let us remind that
 one has to be careful when using $\sigma_{\rm los, 0}^{\rm star}$
or $\sigma_{\rm 3, bo}^{\rm star}$ as estimators for
this physically meaningful quantity.
In Figure~\ref{Cvd4.8} we plot the
$c_{\rm vd} \equiv (\sigma_{\rm 3, h}^{\rm tot} / \sigma_{\rm 3, bo}^{\rm star})^2$
ratios, that measure how dissipation
and concentration affect, on average, to the
relative values of the dispersion at the halo scale (involving also
dark matter) and at the baryonic object  scale. No mass effects are apparent 
in this Figure, but an average kinematical segregation is clear,
(see Table~\ref{TabFP} for the results of a fit to the
expression 
$c_{\rm vd} = A_{\rm vd} (M_{\rm bo}^{\rm star})^{\beta_{\rm vd}}$).  
These are  important results, which could have
interesting observational implications.

\section{The Intrinsic Dynamical Planes and the Dynamical Fundamental Planes
 }
\label{IDPFP}

\subsection{The dynamical plane relations of ELO samples}
\label{DPELOs}

In the last section it has been shown that the
mass, size and velocity dispersion
 parameters {\it at the halo scale} satisfy virial relations.
This result is, however,
at odds with the  tilt of the observed
FP of ellipticals discussed in $\S$\ref{Intro},
that involves the $L$, $R_{\rm e}^{\rm light}$ and $\sigma_{\rm los, 0}$
observed variables, whose virtual counterparts  describe the ELO at
{\it the scale of the baryonic object}.
So, we have first
to analyse whether or not the mass, size and velocity dispersion
of ELOs at this scale define planes tilted relative to the virial one.

To this end, we have carried out 
a principal component analysis (PCA) of the SF-A and SF-B samples
in the  three dimensional variables
$E \equiv \log_{10} M_{\rm bo}^{\rm star}$,
$r \equiv  \log_{10} r_{\rm e, bo}^{\rm star}$ and
$v \equiv \log_{10} \sigma_{\rm 3, bo}^{\rm star}$
through their $3 \times 3$ correlation matrix \textbf{\textsf{C}}.
We have used three dimensional variables rather than projected ones to
circumvent projection effects, that add noise to the correlations
(see, for example, Figure~\ref{Cvpc4.9}).
We have found that,
irrespective of the SF parametrisation,
one of the eigenvalues of \textbf{\textsf{C}} is
considerably smaller than the others (see Table~\ref{PCA}),
so that ELOs populate 
in any case a flattened ellipsoid close to a two-dimensional plane in the
$(E,r,v)$ space that we call the
intrinsic dynamical plane (IDP);
the FP is the observed manifestation of this IDP.
The eigenvectors of \textbf{\textsf{C}} indicate that the projection

\begin{equation}
E - \textit{\~E} = \alpha^{\rm 3D} (r - \textit{\~r}) + \gamma^{\rm 3D} (v - \textit{\~v}),
\label{planeEq}
\end{equation}

where $\textit{\~E}, \textit{\~r}$ and $\textit{\~v}$ are
the mean values of the $E, r$ and $v$ variables,
shows the IDP viewed edge-on.
Table~\ref{PCA} gives the eigenvalues of the correlation matrix
 \textbf{\textsf{C}} ($\lambda_{1}$, $\lambda_{2}$, $\lambda_{3}$), the
planes  eq. (\ref{planeEq}),
as well as their corresponding thicknesses $\sigma_{\it Erv}$,
both for the SF-A and SF-B samples.
The IDPs are in fact tilted relative to the virial
plane (characterized by $\alpha = 1, \gamma = 2$),
and their scatter is very low
as measured by their thicknesses $\sigma_{\it Erv}$.
Note that the eigenvalues of the PCA analysis are not dependend 
on the SF parametrisation.

\begin{table*}
\caption{ Results of PCA at z=0}
\label{PCA}
\begin{center}
\footnotesize
\begin{tabular}{lcccccccccc}
\hline
Sample & No. & $\mbox{\~E}$& $\mbox{\~r}$& $\mbox{\~v}$& $\lambda_{1}$ & $\lambda_{2}$ & $\lambda_{3}$ & $\alpha^{\rm 3D}$& $\gamma^{\rm 3D}$& $\sigma_{\rm Erv}$ \\
\hline
\hline
SF-A& 26& 10.993& 0.746& 2.335& 0.12893 & 0.00355 & 0.00013 & 0.427& 2.066& 0.011\\
SF-B& 17& 11.259& 0.695& 2.453& 0.09342 & 0.00292 & 0.00014 & 0.249& 2.449& 0.012 \\
\hline
\hline

\end{tabular}
\end{center}

\medskip

Column 2: ELO number in the sample.\\
Columns 3, 4 and 5: sample mean values of the $E, r$ and $v$ variables.\\
Columns 6, 7 and 8: eigenvalues of the correlation matrix.\\
Columns 9 and 10: coefficients of the plane (Eq. (5)).\\
Column 11: IDP scatter in the $E, r$ and $v$ variables.

\end{table*}

\begin{figure*}
  \begin{center}
        \includegraphics[width=.45\textwidth]{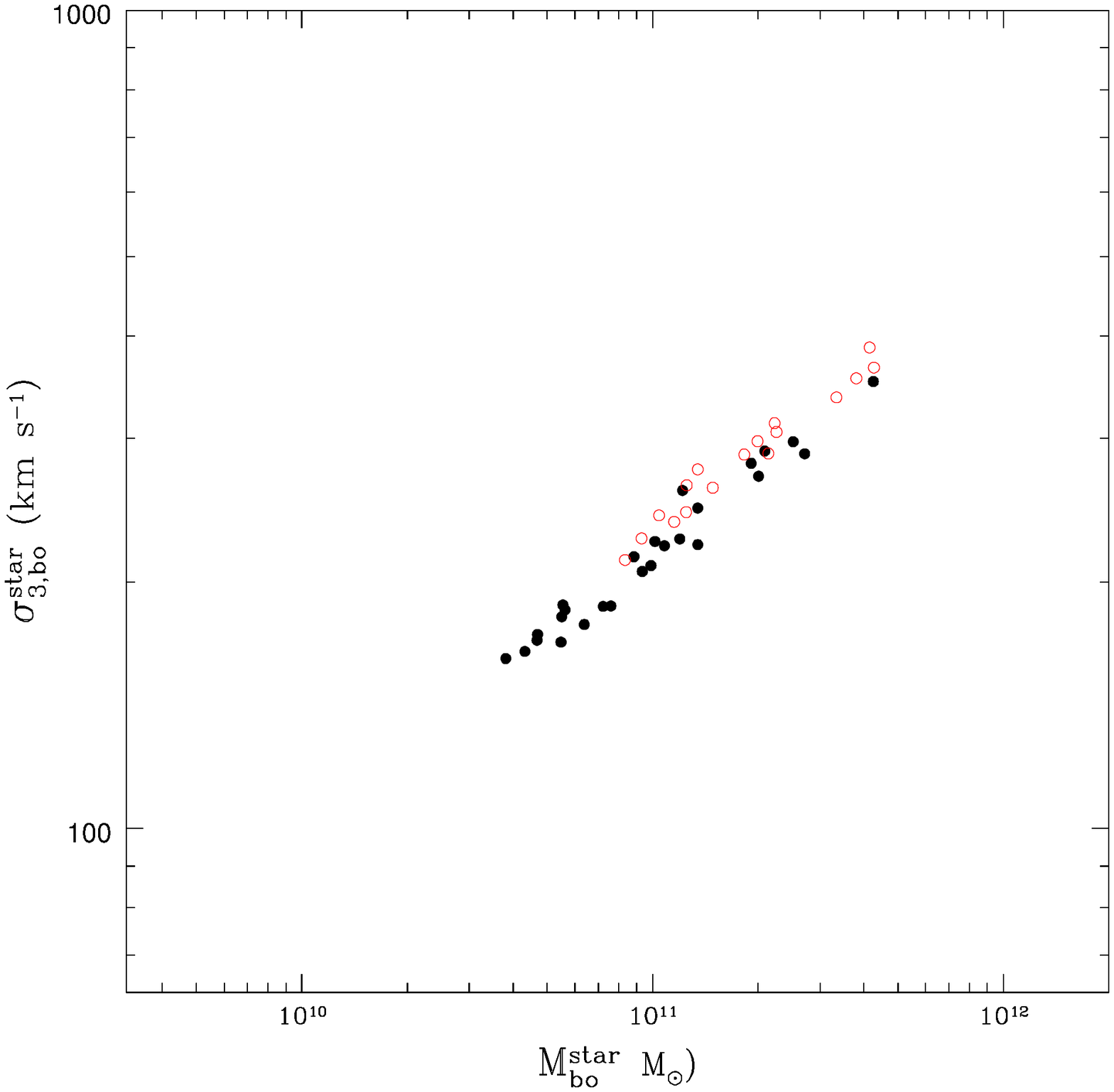}
	\includegraphics[width=.45\textwidth]{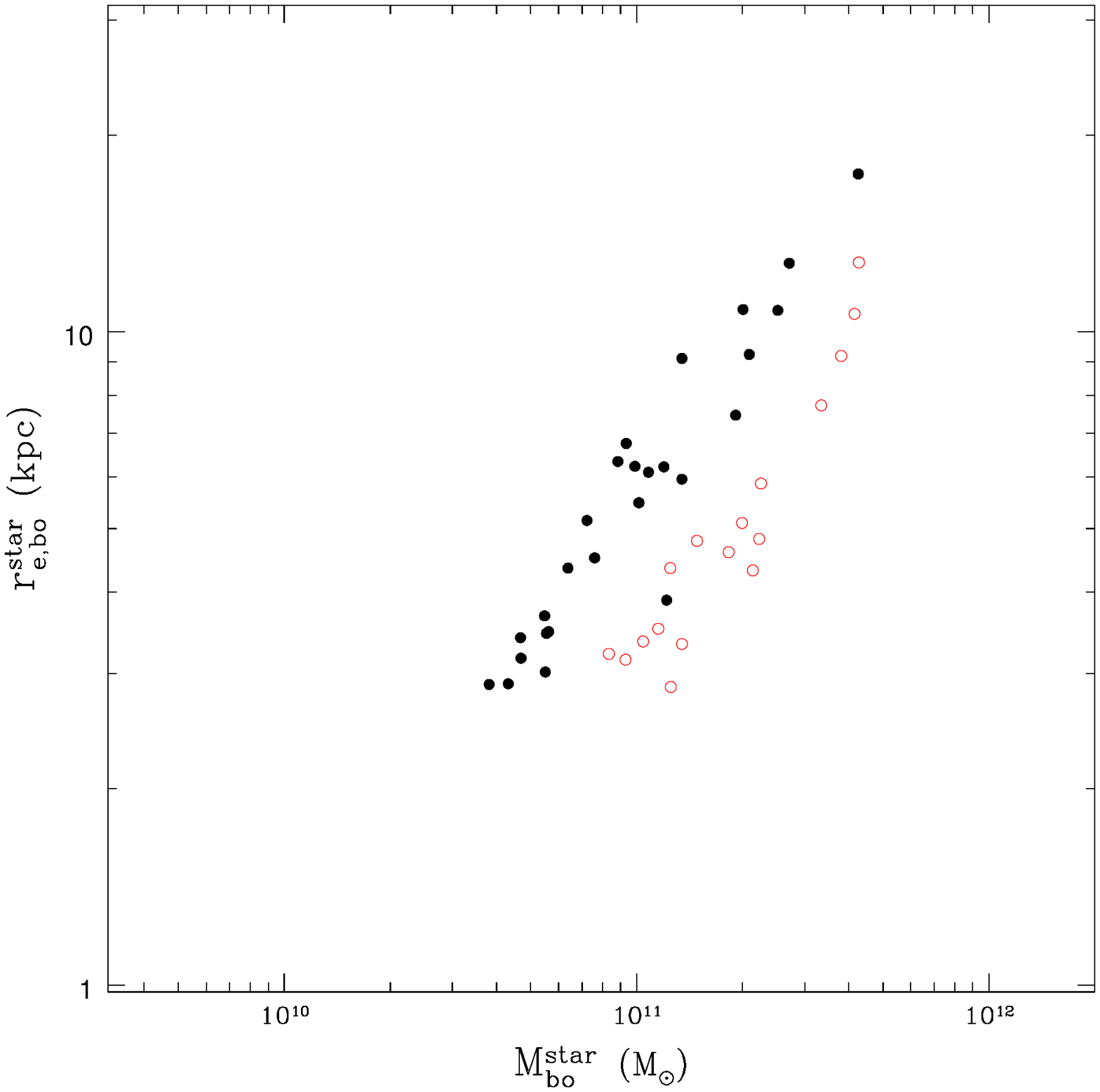}
	\includegraphics[width=.45\textwidth]{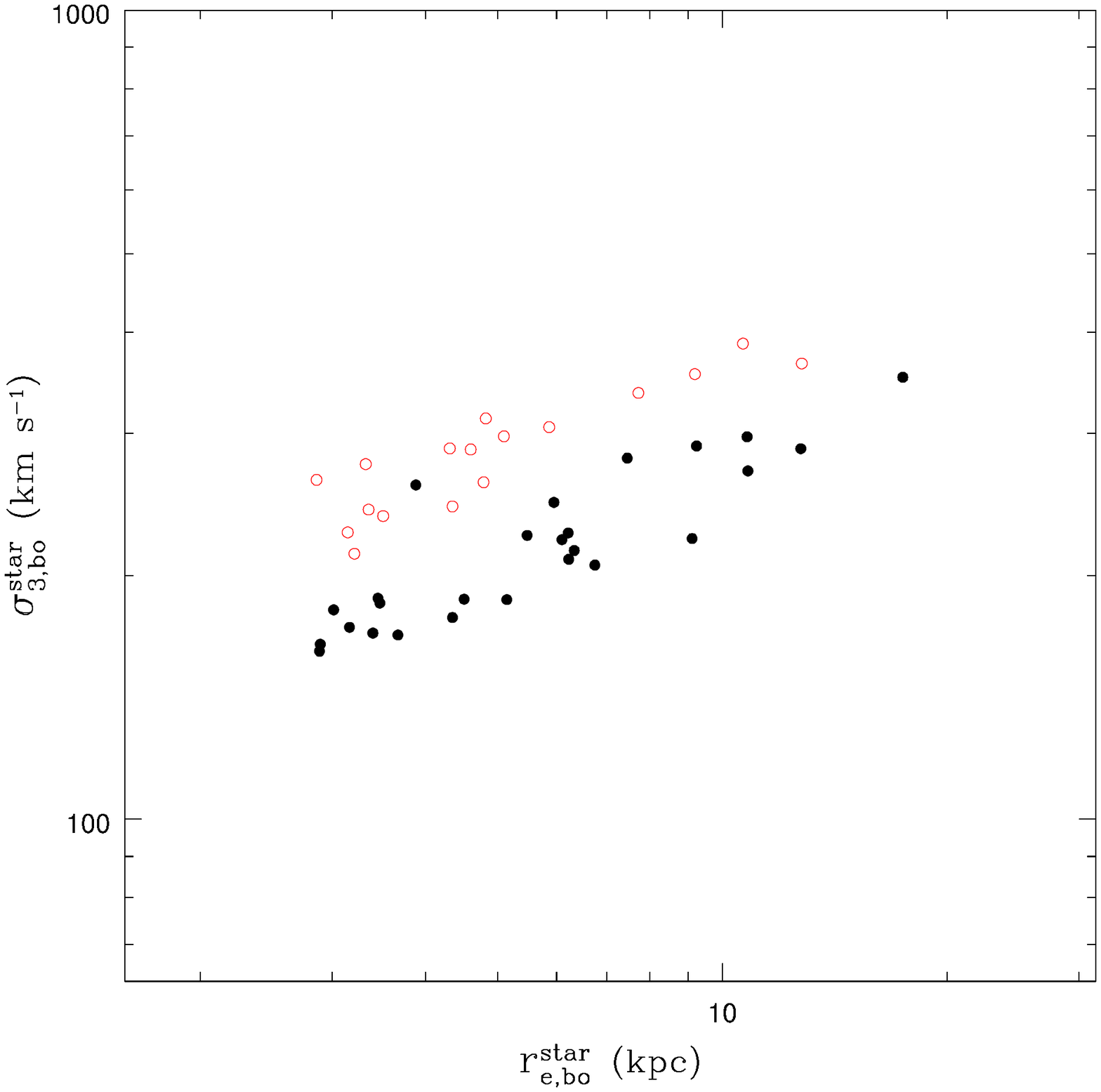}
	          \caption{ The  IDPs for the SF-A and
 SF-B samples. Projections on the $(E, v)$, $(E, r)$ and $(r,v)$
		  coordinate planes are shown. 
		  }
\label{IDynPlane}
\end{center}
\end{figure*}

In Figure~\ref{IDynPlane} we plot the $(E, v), (E, r)$ and $(r,v)$
projections of the IDPs
corresponding  both to the SF-A sample and the SF-B sample.
We see that the three projections show correlations and that
these are very tight for the  first of them.

\subsection{Comparing the IDP to the observed FP of elliptical galaxies}
\label{CompDP}

The next step is to compare the results on the IDPs we have found with
the tilt and the scatter of the
observed FP. We note that the $r$ and $v$ variables are not 
observationally available, so that we have to use their 
projected counterparts.
Assuming that the projected stellar {\it mass} density profile,
$\Sigma^{\rm star}(R)$, can be taken as a measure of the
surface {\it brightness} profile, then
$<\Sigma^{\rm star}>_{\rm e} = c <I^{\rm light}>_{\rm e}$, with $c$ a constant, and
$R_{\rm e, bo}^{\rm star} \simeq R_{\rm e}^{\rm light}$
and we can look for a fundamental plane
(hereafter, the dynamical FP)
in the 3-space of the structural and dynamical parameters
$R_{\rm e, bo}^{\rm star}$, $<\Sigma^{\rm star}>_{\rm e}$ and
$\sigma^{\rm star}_{\rm los, 0}$,
directly provided by the hydrodynamical simulations.
To make this analysis as clear as possible, we transform to
a  $\kappa$-like orthogonal coordinate system,
the dynamical $\kappa_i^{\rm D}$ system, $i$=1,2,3,
 similar to that introduced
 by Bender, Burstein \& Faber (1992), but
 using $R_{\rm e, bo}^{\rm star}$ instead of $R_{\rm e}^{\rm light}$ and
 $<\Sigma^{\rm star}>_{\rm e}$ instead of $<I^{\rm light}>_{\rm e}$,
 and, consequently, free of age,  metallicity  or IMF effects.
The dynamical $\kappa_i^{\rm D}$ variables can be written as:

 \begin{equation}
\kappa_{1}^{D} \equiv [\log \left( \sigma^{\rm star}_{\rm los, 0}\right)^{2}+\log R_{\rm e, bo}^{\rm star}]/\sqrt{2} 
\label{kd1}
\end{equation}

\begin{equation}
\kappa_{2}^{D} \equiv [\log \left( \sigma^{\rm star}_{\rm los, 0}\right)^{2} +2
\log \langle \Sigma^{\rm star} \rangle_{\rm e} - \log R_{\rm e, bo}^{\rm star}]/\sqrt{6}
\label{kd2}
\end{equation}

\begin{equation}
\kappa_{3}^{D} \equiv [\log \left( \sigma^{\rm star}_{\rm los, 0}\right)^{2} - \log \langle \Sigma^{\rm star} \rangle_{e} - \log R_{\rm e, bo}^{\rm star}]/\sqrt{3}
\label{kd3}
\end{equation}

 and they are related  to the $\kappa$ coordinates  through the expressions:
 $\kappa_1 \simeq \kappa^{\rm D}_1$,
 $\kappa_2 \simeq \kappa^{\rm D}_2 - \sqrt 6/3  \log (M_{\rm bo}^{\rm star}/L)$
 and $\kappa_3 \simeq \kappa^{\rm D}_3  +  \sqrt 3/3  \log (M_{\rm bo}^{\rm star}/L)$.
 We discuss the tilt and the scatter of the dynamical FP  separately.
 We first address the tilt issue. We use at this
  stage for the $R_{\rm e, bo}^{\rm star}$ and $\sigma^{\rm star}_{\rm los, 0}$
    variables the averages over three  orthogonal l.o.s.  projections,
    to minimize the scatter in the plots caused by projection
      effects.

\begin{figure*}
  \begin{center}
      \includegraphics[width=.7\textwidth]{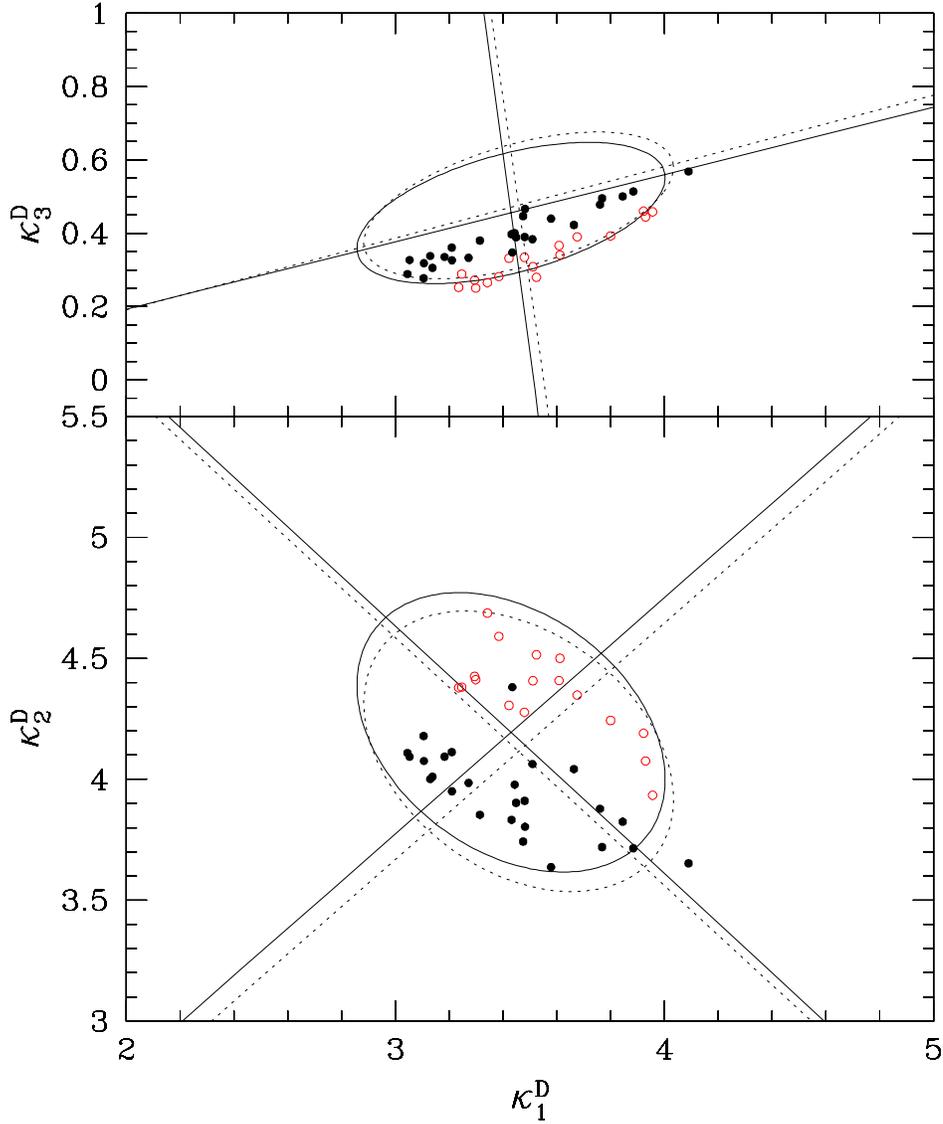}
          \caption{ Dynamical Fundamental Plane in $\kappa^{D}$ system. Edge-on projection (top panel) and nearly
 face-on projection (bottom panel) of the dynamical FP of ELOs in the $\kappa^{D}$ variables (black filled symbols: SF-A sample; red open symbols: SF-B sample). We also draw the respective concentration ellipses (with their major and minor axes) for the SDSS early-type galaxy sample from Bernardi et al. (2003b) in the z band (solid line) and the r band (dashed line). See text for more details.}
\label{kappa}
\end{center}
\end{figure*}

Figure~\ref{kappa} plots the  $\kappa_{3}^{\rm D}$  versus $\kappa_{1}^{\rm D}$
(top) and $\kappa_{2}^{\rm D}$  versus $\kappa_{1}^{\rm D}$ (bottom)
diagrams for ELOs in both the SF-A  and SF-B  samples.
We also drew the 2$\sigma$
concentration ellipses in the respective
variables, as well as its major and minor axes,  for
the SDSS early-type galaxy  sample in the SDSS  $z$ band (solid lines)
and in the $r$
band (point lines) as analysed by
Bernardi et al. 2003b, 2003c.
The most outstanding feature of this Figure
(upper panel) is the good scaling behaviour of
 $\kappa_{3}^{\rm D}$  versus
  $\kappa_{1}^{\rm D}$, with a very low
scatter (see the slopes $M_1$ in Table~\ref{TabFP};
		note that the slopes for the SF-A and SF-B samples are
consistent within their errors, while the zero-points depend on the
SF parameterisation through the ELO sizes).
Another interesting feature of Figure~\ref{kappa}
is that it shows that most of the  values of the $\kappa_{i}^{\rm D}$
coefficients are within the 2$\sigma$ concentration ellipses in both plots for
ELOs formed in SF-A type simulations, with a slightly worse  agreement
for ELOs in the SF-B sample. This means that ELOs have
counterparts in the real world (S\'aiz et al. 2004).
It is worth mentioning that these results  are stable against
slight changes in the values of the $\Omega_{\Lambda}$, $\Omega_{\rm baryon}$
or $h$ parameters;
for example, we have tested that
using their preferred WMAP values (Spergel et al. 2003)
shows results negligibly different to
those plotted in  Figure~\ref{kappa}.
Finally, we note that either the dynamical or the observed FPs are not
homogeneously populated: both SDSS ellipticals and ELOs occupy only
a region within these planes (see  Figure~\ref{kappa} lower panel,
see also Guzm\'an et al. 1993; M\'arquez et al. 2000).
This means that, from the point of view of their
structure and dynamics, ELOs are a two-parameter family where the two
parameters are not fully independent.
Moreover,
concerning ELOs, the occupied region changes when the SF parameters
change.  The reason of this change is that the  ELO sizes decrease
as SF becomes more difficult, because
 the amount of dissipation experienced
by the stellar component along ELO  assembly increases.

We now turn to consider the scatter of the dynamical FP for the ELO samples
and compare it with the scatter of the FP for the SDSS elliptical sample,
calculated as the square root of the smallest eigenvalue of the 3$\times$3
covariance matrix in the $E$ (or $\log L$),
$V \equiv \log \sigma_{\rm los, 0}^{\rm star}$ and 
$R \equiv \log R_{\rm e, bo}^{\rm star}$ variables (Saglia et al. 2001).
As Figure~\ref{kappa} suggests,
when projection effects are circumvented by taking averages
over different directions, the resulting three dimensional orthogonal scatter
for ELOs is smaller than for SDSS ellipticals ($\sigma_{\it EVR} = 0.0164$
and $\sigma_{\it EVR} = 0.0167$ for the SF-A and SF-B samples,
respectively, to be compared with
$\sigma_{\it LVR} = 0.0489$  for the SDSS
in the $\log L, V \equiv \log \sigma_{\rm los, 0} $ and 
$R \equiv \log R_{\rm e}^{\rm light}$ variables).
To estimate the contribution of projection
effects to the observed scatter, we have calculated the  orthogonal
scatter for ELOs when no averages over projection directions
for the $R_{\rm e, bo}^{\rm star}$ and $\sigma_{\rm los, 0}^{\rm star}$
variables are made. The scatter ($\sigma_{\it EVR} = 0.0238 $
and $\sigma_{\it EVR} = 0.0214$ for the SF-A and SF-B samples)
increases, but it is still
lower than observed. This indicates  that a contribution
from stellar population effects is needed to explain
 the scatter of the observed FP, as suggested by  different authors
 (see, for example,  Pahre et al. 1998; Trujillo et al. 2004).

\subsection{Clues on the physical origin of the IDP tilt}
\label{CluesTilt}

We now address the issue of the physical origin of the tilt
of ELO IDPs relative to the virial relation.
As discussed in $\S$\ref{Intro},
a non-zero tilt can be caused by a mass dependence of the mass-to-light
ratio $M_{\rm vir}/L$, of the mass structure coefficients
$c_{\rm M}^{\rm vir} \equiv {G M_{\rm vir} \over 3 \sigma_{\rm los, 0}^{2} R_{\rm e}^{\rm light}}$, or of both of them. We examine these possibilities in turn.

i) We first note that the  mass-to-light ratio can be written as:

\begin{equation}
M_{\rm vir}/L = A_{\rm vir} (M_{\rm bo}^{\rm star})^{\beta_{\rm vir}} \times \gamma^{\rm star},
\end{equation}

where $\gamma^{\rm star} \equiv M_{\rm bo}^{\rm star}/L$ 
is the stellar mass-to-light ratio, that, as already explained,
can be considered to be independent of the E galaxy luminosity
or ELO mass scale.
Figure~\ref{CMvirMob4.5} and the  values of the $\beta_{\rm vir}$ slopes
given in Table~\ref{TabFP}, indicate that the dark to bright mass content
of ELOs increases with their mass, contributing a tilt to their IDPs. Similar
results have also been found in pre-prepared simulations of
dissipative mergers (Robertson et al. 2006).

ii) Writting the  $c_{\rm M}^{\rm vir}$ mass structure coefficients 
as power laws
$c_{\rm M}^{\rm vir} = A_{\rm M} (M_{\rm bo}^{\rm star})^{\beta_{\rm M}}$,
ELO homology would imply ${\beta_{\rm M}} = 0$.
To elucidate whether or not this is the case, the 
${\beta_{\rm M}}$ slopes have been measured on  the ELO samples
through direct fits in log-log scales. The results are given in
Table~\ref{TabFP}, where we see that the homology is in fact broken
both for SF-A or SF-B samples.
To deepen into the causes of this behaviour,
we use Eqs. \ref{Virial} and \ref{CMvir}
to write 

\begin{equation}
c_{\rm M}^{\rm vir} = c_{\rm F} c_{\rm rd} c_{\rm rp} c_{\rm vd} c_{\rm vpc},
\end{equation}

where the $c_{\rm i}$ coefficients, with i = rd, rp, vd, vpc
have been defined in $\S$\ref{ResBarObj}, and with i = F
in $\S$\ref{GloLOSVDPar}. Taking into account the power-law forms
of these coefficients, we have:

\begin{equation}
\beta_{\rm M} = \beta_{\rm F} + \beta_{\rm rd} + \beta_{\rm rp} + \beta_{\rm vd} + \beta_{\rm vpc},
\end{equation}

when the $\beta_{\rm i}$ slopes are calculated
through direct fits. From Table~\ref{TabFP}
we see that
the main contribution to the homology breaking comes from
the $c_{\rm rd}$ coefficients (i.e., spatial homology breaking, see
$\S$\ref{ResBarObj} and O\~norbe et al. 2005),
while, as previously discussed, no dynamical
homology breaking (see $\S$\ref{GloLOSVDPar})
or projection effects are important in our ELO samples.

\section{Summary, discussion  and conclusions}

\subsection{Summary}
\label{summary}
We present an analysis of the sample of ELOs formed in ten different
cosmological  simulations, 
run within the same
global flat $\Lambda$ cosmological model, roughly consistent
with observations.
The normalisation parameter has been taken slightly high,
$\sigma_8 = 1.18$, as compared with the average fluctuations
of 2dFGRS or SDSS  galaxies
to mimic an active region of the Universe.
Newton laws and hydrodynamical
 equations have been integrated in this context,
 with a standard cooling algorithm and a star formation parameterization
 through a Kennicutt-Schmidt-like law,
 containing our ignorance about its details at sub-kpc scales.
No further hypotheses to model  the assembly processes have been made.
 Individual galaxy-like objects
  naturally appear as an output of the simulations, so that
   the physical processes underlying mass assembly can be studied.
Five out of the ten simulations 
 (the SF-A type simulations)
 share the SF parameters 
 and differ in the seed  used to build up the initial conditions.
 To test the role of SF parameterisation,
 the same initial conditions
 have been run with different SF parameters
 making SF  more difficult, contributing another set of five
 simulations  (the SF-B type simulations).
 ELOs have been identified in the simulations as those
 galaxy-like objects  at $z = 0$ having a prominent, non-rotating
 dynamically relaxed spheroidal   component made out
 of stars, with no extended discs and very low gas content.
 These stellar component is embedded in  a dark matter halo
 that contributes an important fraction
  of the mass at distances from the ELO centre larger than
  $\sim 10 - 15$ kpc. 
  No ELOs with stellar masses below 3.8 $\times$ 10$^{10}$ M$_{\odot}$
   or virial masses below 3.7 $\times$ 10$^{11}$ M$_{\odot}$ have been
    found that met the selection  criteria
     (see Kauffmann et al. 2003b for a similar result in SDSS galaxies
      and Dekel \& Birnboim 2006, and Cattaneo et al. 2006 for a possible
 physical explanation).
 ELOs have also an extended halo of hot, diffuse gas.
  Stellar and dark matter particles constitute a dynamically hot
  component with an important velocity dispersion, and, except
  in the very central regions, a positive anisotropy.

The informations about position and velocity distributions
of the ELO   particles of different kinds (dark matter, stars, cold gas, hot gas)
provided by the simulations, allows a
detailed study of
their  intrinsical three dimensional
mass and  velocity distributions, as well as a measure of the parameters
characterising their structure and dynamics. 
In a forthcoming paper, we report on the three dimensional mass
density, circular velocity and velocity dispersion  profiles,
as well as the projected stellar mass density profiles and the LOS velocity
dispersion profiles.  
In this paper we focus on a  parameter analysis to quantify
some of the previous results.

Mass, size and velocity dispersion scales for their different
components have been measured in the ELO samples,
both at the scale of their halo and at the scale of the baryonic object
(a few tens of kiloparsecs).
At the {\it halo scale}, the masses of both cold gas and stars,
$M_{\rm h}^{\rm cb}$ and $M_{\rm h}^{\rm star}$, respectively, have been
found to be tightly correlated with the halo total mass,
$M_{\rm vir}$, with the ratios $M_{\rm h}^{\rm cb}/M_{\rm vir}$
and $M_{\rm h}^{\rm star}/M_{\rm vir}$ decreasing as $M_{\rm vir}$
increases (that is, massive objects miss cold baryons within
$r_{\rm vir}$ when compared with less massive ELOs),
presumably because gas
gets more difficulties to cool and fall as $M_{\rm vir}$ increases.
The overall half-mass radii, $r_{\rm e, h}^{\rm tot}$ shows also a
very tight correlation with $M_{\rm vir}$.
Half-mass radii for the cold baryon or stellar mass distributions
have a more complex behaviour, as in these cases
gas heating in  shocks and energy losses due gas cooling 
are in competition to determine
these distributions.
 
A very important result
we have found when analysing ELOs
at the {\it scale of the baryonic object},
is that $M_{\rm vir}$ plays an important role  
to determine the ELO structure also below a few tens of kiloparsecs scales.
In fact, both the masses of cold baryons $M_{\rm bo}^{\rm cb}$
(i.e., those baryons that have reached the
central regions of the configuration), and of stars $M_{\rm bo}^{\rm star}$,
show a good correlation with $M_{\rm vir}$, and, moreover,
the  $M_{\rm bo}^{\rm cb}/M_{\rm vir}$ and
$M_{\rm bo}^{\rm star}/M_{\rm vir}$ ratios
(i.e., the relative content of cold baryons or stars versus
total mass) decrease as $M_{\rm vir}$
increases. This is the same qualitative  behaviour shown by these ratios
observationally in the SDSS data, and, also, by ELOs at the halo scale.
The dependence of $M_{\rm bo}^{\rm cb}$ or $M_{\rm bo}^{\rm star}$
on the SF parametrisation is only very slight, with SF-A type ELOs having
slightly more stars than their SF-B type counterparts.
The half-mass radii for cold baryon and stellar masses,
$r_{\rm e, bo}^{\rm cb}$ and $r_{\rm e, bo}^{\rm star}$,
show also a good correlation with $M_{\rm vir}$, but now the values of the
SF parameters also play a role, because
their change implies a change
in the time interval during which gas cooling is turned on,
and this changes  the ELO  stellar
mass distribution, i.e., its lenghthscale,
 so that ELO compactness
increases from SF-A to SF-B type simulations.
Another important result is that,
regardless of the SF parametrizations used in this work,
the relative distributions of the
stellar and dark mass components in ELOs show a systematic trend
measured through the 
$c_{\rm rd} \equiv r_{\rm e, h}^{\rm tot} / r_{\rm e, bo}^{\rm star}$ ratios,
with stars relatively more concentrated as $M_{\rm vir}$ decreases
(i.e., a quantification of the spatial  homology breaking).
Note that to compare with observational data,
the relevant parameters are the
{\it projected} half-mass radii,
$R_{\rm e, bo}^{\rm star}$. We have checked that they
show an excellent correlation with the corresponding three dimensional
half-mass radii, with the
 $c_{\rm rp} \equiv r_{\rm e, bo}^{\rm star} / R_{\rm e, bo}^{\rm star}$
 ratios showing no significant dependence on the ELO mass scale.

Concerning kinematics, a useful characterisation of the ELO velocity dispersion
is the central stellar line-of-sight velocity dispersion,
$\sigma_{\rm los, 0}^{\rm star}$, whose observational
counterpart can be measured from
elliptical spectra. A very important result is the very tight
correlation we have found between $M_{\rm vir}$ and
$\sigma_{\rm los, 0}^{\rm star}$, confirming that
the observationally measurable 
$\sigma_{\rm los, 0}$ is a fair virial mass estimator.
In addition, $\sigma_{\rm los, 0}^{\rm star}$
is closely related to the mean square velocity of
both, the whole elliptical at the halo scale (including the dark matter),
$\sigma_{\rm 3, h}^{\rm tot}$, and the stellar component of the central object, 
$\sigma_{\rm 3, bo}^{\rm star}$. We have also found that  the 
$c_{\rm vd} \equiv (\sigma_{\rm 3, h}^{\rm tot} / \sigma_{\rm 3, bo}^{\rm star})^2$ or the 
$c_{\rm vpc} \equiv (\sigma_{\rm 3, bo}^{\rm star})^2 / 3 (\sigma_{\rm los, 0}^{\rm star})^2$
ratios are roughly independent of the ELO mass scale.
And so, ELOs do not show dynamically broken homology, even if
their stellar and
dark components are kinematically segregated (i.e., $c_{\rm vd} \ne 1$).
This could lead to inaccurate determinations of the total
mass of ellipticals when using stellar kinematics.

A very important result is that, irrespective of the SF parameterisation,
the (logarithms of the)
ELO stellar  masses $M_{\rm bo}^{\rm star}$,
 stellar half-mass radii  $r_{\rm e, bo}^{\rm star}$, and stellar 
 mean square velocity of the central object $\sigma_{\rm 3, bo}^{\rm star}$,
 define {\it intrinsic dynamical}  planes (IDPs).
 These planes are tilted relative to the virial plane and the tilt
 does not significantly depend on the SF parameterisation, but the
 zero point does depend.
 Otherwise, the intrinsic dynamical plane is not homogeneously
 populated, but ELOs, as well as E galaxies in the FP
 (Guzm\'an et al. 1993), occupy only a particular region defined by
 the range of their masses.

\subsection{Testing possible resolution effects}
\label{resoleff}
\begin{figure}
  \begin{center}
    \includegraphics[width=.45\textwidth]{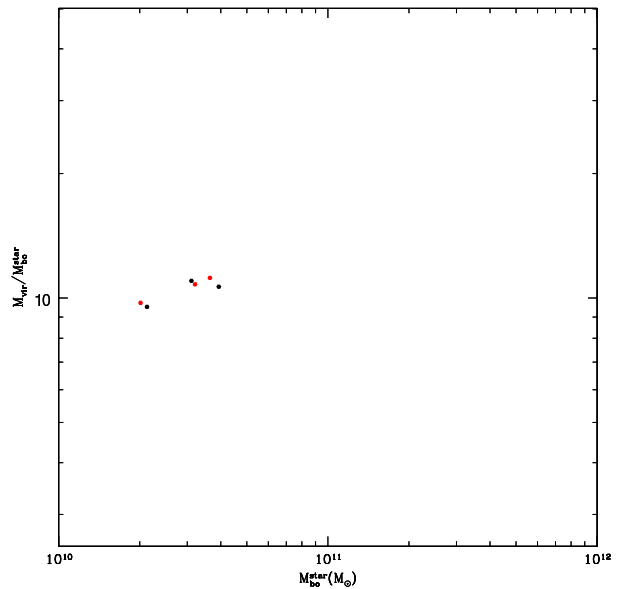}
\caption
{The $M_{\rm h}^{\rm cb}/M_{\rm bo}^{\rm cb}$ ratios as a function of
      the ELO mass scale for three ELOs identified at
$z=1$ in S128 (red) and their counterparts in S64 (black).
}
   \label{ComCoc}
  \end{center}
\end{figure}

To make sure that the results we report in this paper are not unstable under
resolution changes, a control simulation with 128$^3$ dark matter and
128$^3$ baryonic particles, a gravitational softening of $\epsilon = 1.15$ kpc
and the other parameters as in SF-A type simulations (the S128 simulation),
has been run. The results of its analysis have been compared with those of a
2 $\times 64^3$ simulation (the S64 simulation),
whose initial conditions have been built up
by randomly choosing 1 out of 8 particles in the S128 initial conditions,
so that every ELO in S128 has a counterpart in the lower resolution
simulation and conversely. Due to the very high CPU time requirements
for S128, the comparison has been made at $z=1$.
The results of this comparison are very
satisfactory, as Figure~\ref{ComCoc} illustrates
(compare with Figure~\ref{CMvirMob4.5}).

\subsection{The dimensionality of ELO and elliptical samples in parameter space}
\label{dimensiomality}
The intrinsic dynamical planes and their occupations (see \S ~\ref{summary})
reflect the fact that dark matter haloes are
 a two-parameter family (for example, the virial mass and the energy content
 or the concentration; see, for example,
 Hernquist 1990;
 Navarro, Frenk \& White 1995, 1996;
  Manrique et al. 2003; Navarro et al. 2004)
 where the two parameters are correlated  (see, for example,
Bullock et al. 2001; Wechsler et al. 2002;  Manrique et al. 2003).
 Adding gas implies that heating and cooling processes also
 play a role at determining the mass and velocity
 distributions, and, more particularly, the length scales.
 However, as explained above,
 we have found that both, 
 the relative content and the relative distributions of the
 dark and baryonic mass components
show systematic trends with the ELO mass scale,
that can be written as power-laws of the form
$ M_{\rm vir}/M_{\rm bo}^{\rm star} = A_{\rm vir} (M_{\rm bo}^{\rm star})^{\beta_{\rm vir}}$ and 
$ r_{\rm e, h}^{\rm tot} / r_{\rm e, bo}^{\rm star} = A_{\rm rd} (M_{\rm bo}^{\rm star})^{\beta_{\rm rd}}$.
     
  A first consequence of the regularity of the trends
  with the mass scale found in this paper is that no new parameters are
  added relative to the dark matter halo family,
  so that  the baryonic objects are also a two-parameter family,
  and ELO structural and dynamical parameters define also a plane.
  A second consequence is that the plane is tilted relative
  to the halo plane (i.e., the virial plane) 
  because $\beta_{\rm rd} - \beta_{\rm vir} \neq 0$.
Finally, the plane is not homogeneously populated because of
the mass-concentration halo correlation, that
at the scale of the baryonic objects appears for example
as a mass---size correlation.
This explains the role played by $M_{\rm vir}$ to
determine the intrinsic three dimensional correlations.
In this paper  we also show that $\sigma_{\rm los, 0}^{\rm star}$
is a fair empirical estimator of $M_{\rm vir}$,
and this explains the central role
played by $\sigma_{\rm los, 0}$ at determining the observational
correlations.

 The fundamental plane shown by real elliptical samples is the
 observationally manifestation of the IDPs when using projected
 parameters $R_{\rm e, bo}^{\rm star}$, $\sigma_{\rm los, 0}^{\rm star}$
 and luminosity variables instead of stellar masses $M_{\rm bo}^{\rm star}$.
 We have taken advantage of the constancy of the stellar-mass-to-light ratios
 of ellipticals in the SDSS (Kauffmann et al. 2003a, 2003b)
 to put the elliptical sample of Bernardi et al. (2003b, 2003c) in the same
 projected variables we can measure in our virtual ellipticals. 
 We have found that the FPs shown by the two ELO samples
 are consistent with that shown by the SDSS elliptical sample in the
 same variables, with
  no further need for any relevant contribution
  from stellar population effects to explain the observed tilt.
  These effects could, however, have contributed to the scatter of the
  observed FP, as the IDPs have been found to be thinner
  than the observed FP.

\subsection{The physical origin of the tilt in a cosmological context}
\label{physorigtilt}
We now turn to discuss  the physical origin of
the trends given by  the power laws
$ M_{\rm vir}/M_{\rm bo}^{\rm star} = A_{\rm vir} (M_{\rm bo}^{\rm star})^{\beta_{\rm vir}}$ and
$ r_{\rm e, h}^{\rm tot} / r_{\rm e, bo}^{\rm star} = A_{\rm rd} (M_{\rm bo}^{\rm star})^{\beta_{\rm rd}}$.
As explained in Section 2, the simulations provide us with clues on
the physical processes involved in elliptical formation
(see also DSS04 and DTal06).
They also  indicate that the dynamical plane
appears at an early violent phase as a consequence
of ELO assembly out of gaseous material, with cooling  and on short timescales,
and it is preserved  during a later, slower phase,
where dissipationless merging plays an important role in
 stellar mass assembly (see more details in DTal06).
 Our simulations show  that the physical origin of the trends
 above lie in the systematic decrease,
 with increasing ELO mass,
 of the relative amount of
 dissipation experienced by the baryonic mass component along ELO
 stellar mass assembly (DTal06; O\~norbe et al., in preparation).
This possibility had been suggested by Bender et al. (1992),
Guzm\'an et al. (1993) and Ciotti et al. (1996).
Bekki (1998) first addressed it numerically in the framework of the
merger hypothesis for elliptical formation through pre-prepared simulations
of dissipative mergers of disk galaxies,
where the rapidity of the star formation in mergers is controlled by
a free efficiency parameter $C_{\rm SF}$. He shows that the star formation
rate history of galaxies determine the differences in dissipative
dynamics, so that to explain the slope of the FP he {\it needs to assume}
that more luminous galaxies are formed by galaxy mergers with a shorter
timescale for gas transformation into stars.
Recently, Robertson et al. (2006) have confirmed these findings
on the importance of dissipation to explain the FP tilt. 

In this paper we go an step further and analyze the FP
of virtual ellipticals formed in a  cosmological context,
where individual galaxy-like objects
 naturally appear as an output of the simulations.
 Our results essentially include previous ones and add
 important new informations.
First, our results on the role of dissipation to produce the
tilt of the FP essentially agree with those obtained through
dissipative pre-prepared mergers, but it is important
to note that, moreover,
more massive objects produced in the
simulations {\it do have} older  means  and narrower spreads
in their stellar age distributions than less massive ones
(see details DSS04); this naturally appears in the
simulations and need not be considered as an additional
assumption.
Second, the preservation of the FP in the slow phase of mass aggregation
in our simulations also
agrees with previous work based on dissipationless simulations
of pre-prepared mergers (Capelato et al. 1995; Dantas et al. 2003;
Goz\'alez-Garc{\'\i}a \& van Albada 2003;  Boylan-Kolchin et al. 2005; 
Nipoti, Londrillo \& Ciotti 2003).
 Moreover, elliptical properties recently inferred from observations
 (for example, the appearence of
 blue cores,  Menanteau et al. 2004, and the increase of the stellar mass
 contributed by the elliptical population since higher $z$,
 Bell et al. 2004; Conselice, Blackburne, {\&} Papovich 2005;
 Faber et al. 2005;
 see more details in DTal06) can also be explained in our simulations.

\subsection{Conclusions} 
\label{conclusion}

 We conclude that the
 simulations provide an unified scenario where most current
 observations on ellipticals can be interrelated. In particular, it explains the most important
results relative to the physical origin of their FP relation
(i.e. the FP tilt is due dissipative dynamics, and disipationless
merging in the slow growth phase of mass assembly does not
change the tilt). It is worth mentioning that this scenario
 shares some characteristics of previously proposed scenarios,
 but it has also significant differences, mainly
 that most stars in elliptical galaxies form out of cold gas that had never been shock heated
 at the halo virial temperature and then formed a disk,
 as the conventional recipe for galaxy formation propounds
 (see discussion in Keres et al. 2005 and references therein).
 The scenario for elliptical formation emerging from our simulations
 has the advantage that it results from simple
 physical laws acting on initial conditions that are
 realizations of power spectra consistent
 with observations of CMB anisotropies.

This work was partially supported by the MCyT (Spain) through grants
AYA-0973, AYA-07468-C03-02 and AYA-07468-C03-03
  from the PNAyA, and also by the regional government of Madrid 
through the ASTROCAM Astrophysics network.
 We  thank the Centro de Computaci\'on
  Cient\'{\i}fica (UAM, Spain) for computing facilities.
  AS  thanks  FEDER financial support from Eropean Union.

\end{document}